\documentclass[reprint, aps, prl, floatfix, superscriptaddress, longbibliography]{revtex4-1}
\usepackage{graphicx}    
\usepackage{dcolumn}    
\usepackage{bm}             
\usepackage{amssymb}   
\usepackage{amsmath}    
\usepackage{amsthm}
\usepackage{mathrsfs}    
\usepackage{commath}   
\usepackage{subfigure}    
\usepackage{braket}
\usepackage{float}
\usepackage{color}
\usepackage{siunitx}
\usepackage{mathtools}
\usepackage{multirow}
\usepackage[colorlinks=true, allcolors=blue]{hyperref}
\usepackage{hyphenat}

\newcommand{\beq}{\begin{equation}}
\newcommand{\eeq}{\end{equation}}

\begin{document}

\title{Quantum Microwave Radiometry with a Superconducting Qubit}

\author{Zhixin~Wang}
\email{zhixin.wang@yale.edu}
\affiliation{Department of Applied Physics and Physics, Yale University, New Haven, CT 06520, USA}

\author{Mingrui~Xu}
\affiliation{Department of Electrical Engineering, Yale University, New Haven, CT 06520, USA}

\author{Xu~Han}
\affiliation{Department of Electrical Engineering, Yale University, New Haven, CT 06520, USA}

\author{Wei~Fu}
\affiliation{Department of Electrical Engineering, Yale University, New Haven, CT 06520, USA}

\author{Shruti~Puri}
\affiliation{Department of Applied Physics and Physics, Yale University, New Haven, CT 06520, USA}

\author{S.~M.~Girvin}
\affiliation{Department of Applied Physics and Physics, Yale University, New Haven, CT 06520, USA}

\author{Hong~X.~Tang}
\affiliation{Department of Electrical Engineering, Yale University, New Haven, CT 06520, USA}

\author{S.~Shankar}
\altaffiliation{Present address: Department of Electrical and Computer Engineering, University of Texas, Austin, TX 78758, USA}
\affiliation{Department of Applied Physics and Physics, Yale University, New Haven, CT 06520, USA}

\author{M.~H.~Devoret}
\email{michel.devoret@yale.edu}
\affiliation{Department of Applied Physics and Physics, Yale University, New Haven, CT 06520, USA}

\date{\today}

\begin{abstract}
The interaction of photons and coherent quantum systems can be employed to detect electromagnetic radiation with remarkable sensitivity.
We introduce a quantum radiometer based on the photon-induced-dephasing process of a superconducting qubit for sensing microwave radiation at the sub-unit-photon level.
Using this radiometer, we demonstrated the radiative cooling of a 1-K microwave resonator and measured its mode temperature with an uncertainty $\sim 0.01\ \si{K}$.
We have thus developed a precise tool for studying the thermodynamics of quantum microwave circuits, which provides new solutions for calibrating hybrid quantum systems and detecting candidate particles for dark matter.
\end{abstract}

\maketitle

Historically, radiometry experiments---measurements of the intensity of radiation---revealed the first invisible band of the electromagnetic spectrum \cite{Herschel1800}, unveiled the law of radiative energy transfer \cite{Dulong1818, Tyndall1865a}, provoked the genesis of quantum theory \cite{Lummer1900, Rubens1900}, and provided the landmark evidence for the Big Bang model of the universe \cite{Penzias1965, Roll1966, Howell1966}.
Contemporarily, a number of investigations in mesoscopic \cite{Blanter2000} and particle physics \cite{Grahan2015} are demanding microwave radiometry with single-photon sensitivity.
In this article, we present a quantum-limited radiometry protocol based on the photon-induced dephasing of a superconducting qubit.
By converting a small photon population to qubit coherence times, this mechanism is particularly suited for detecting microwave radiation beyond the Rayleigh--Jeans regime where single-photon energy exceeds the energy of thermal fluctuations.

A classical microwave radiometer \cite{Skou2006}
typically consists of a radio receiver followed by a square-law detector that converts the input power to an output electrical quantity.
In reality, receiving circuits always add noise to the antenna radiation.
This system noise, represented by a \emph{system temperature} $T_\mathrm{sys}$, limits the sensitivity of a radiometer \cite{sensitivity}.
A common practice for reducing $T_\mathrm{sys}$ is to equip the receiver with cryogenic low-noise amplifiers \cite{Bryerton2013}.
Following this design, space-based radiometers with $T_\mathrm{sys} \sim 10 \textrm{--} 100\ \si{K}$ have been developed to map the anisotropies of the cosmic microwave background radiation \cite{Jarosik2003,Aja2005}.
Alternatively, one can directly detect quantum radiation with novel microwave sensors.
For instance, in circuit quantum electrodynamics (cQED) systems \cite{Blais2004, Blais2007, Schoelkopf2008, Girvin2014},
one can deduce the coherent amplitude \cite{Schuster2005} or white-thermal-photon population \cite{Bertet2005a,Rigetti2012,Sears2012,Yan2016,Yan2018,Yeh2017,Wang2019} of a microwave cavity from qubit coherence times---an effect known as photon-induced qubit dephasing \cite{Bertet2005b,Gambetta2006,Clerk2007}.
This method has been applied to extract the residual thermal populations of qubit-readout cavities with the highest reported precision $\sim10^{-4}$ \cite{Wang2019}.
However, qubit radiometers have not yet been employed to probe radiative sources with pulsed emission or colored spectra.

A frequently encountered type of thermal radiator is an electrical resonator at finite temperature, whose output field has a Lorentzian power spectral density centered at its resonant frequency.
Consider a resonator mode at frequency $f$ coupled with rate $\kappa_\mathrm{c}$ to a transmission line at temperature $T_\mathrm{c}$, and with rate $\kappa_\mathrm{i}$ to an internal dissipative mechanism thermalized to a bath at $T_\mathrm{i} > T_\mathrm{c}$. The average thermal population of this resonator mode is $\bar{n} = (\kappa_\mathrm{c} \bar{n}_\mathrm{c} + \kappa_\mathrm{i} \bar{n}_\mathrm{i})/(\kappa_\mathrm{c}+\kappa_\mathrm{i})$, in which $\bar{n}_\mathrm{c(i)} = (e^{h f / k_\mathrm{B} T_\mathrm{c(i)}}-1 )^{-1}$.
\emph{Radiative cooling} ($\bar{n} \rightarrow \bar{n}_\mathrm{c}$) is manifested if $\kappa_\mathrm{c} \gg \kappa_\mathrm{i}$, namely, when the resonator is over-coupled to a colder external bath.
The same thermodynamic principle underlies nocturnal cooling on cloudless nights \cite{Brunt1932,Groen1947} when the earth's surface, in contact with the 300-K atmosphere, is cooled by the 3-K cosmic microwave background.
In this work, to test our qubit radiometer, we measured the thermal radiation from a 1-K microwave resonator, and unambiguously demonstrated its radiative cooling by a 0.1-K external coupling bath.
Noise analysis yields $T_\mathrm{sys}=0.31\ \si{\K} \sim hf/2k_\mathrm{B}$, which is dominated by dissipation and thermal leakage at low temperatures, and is a factor of two below the minimum output noise energy $hf$ of an ideal quantum-limited phase-preserving linear amplifier, which includes both the amplified vacuum fluctuations accompanying the input signal and the minimum possible noise added by the amplifier.
The well-understood performance and the exceptional precision of this radiometer promise its applicability to a variety of problems, such as verifying ground-state cooling in hybrid quantum systems \cite{Zhong2019,spin_cool} and upgrading microwave-cavity-based axion experiments \cite{Lamoreaux2013,Zheng2016,Dixit2018}.

\section{Measurement setup}

\begin{figure} [t]
 \includegraphics[angle = 0, width = \columnwidth]{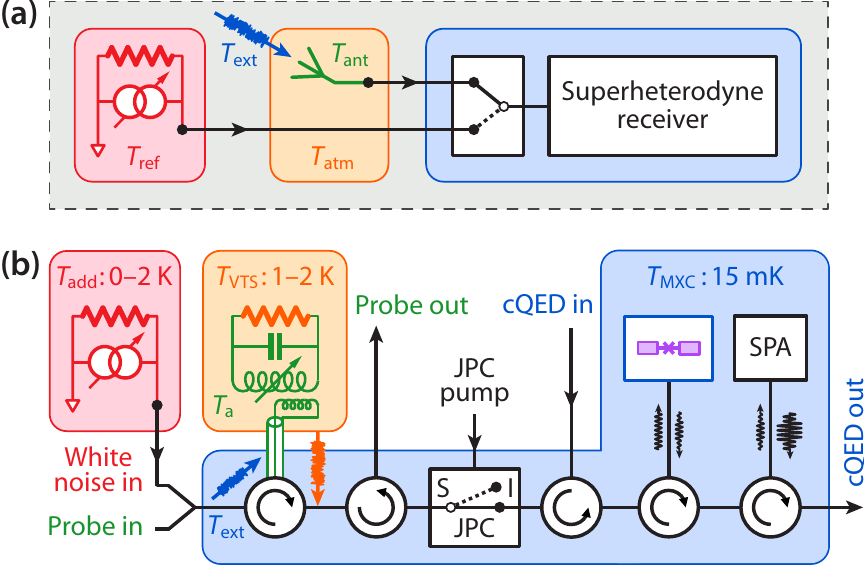}
 \caption{
 \textbf{(a)}
 Diagram of an archetypal radio telescope.
 Blue box: radio receiver.
 Orange box: earth's atmosphere.
 Green trident: receiving antenna.
 Blue arrow: extraterrestrial radiation.
 Red box: reference black-body radiator for temperature calibration.
 \textbf{(b)}
 Setup of our quantum microwave radiometric experiment.
 Blue area: mixing-chamber stage (MXC) of a dilution refrigerator.
 Orange box: variable-temperature stage (VTS), which is attached to the still stage of the same dilution fridge through a weak thermal link.
 Green transmission line: a niobium--titanium superconducting coaxial cable inductively coupled to a tunable antenna resonator (green).
 Blue arrow: residual thermal noise from the input line impinging on the resonator.
 Orange arrow: thermal leakage from the VTS to the MXC circuit.
 Red box: tunable broadband noise generator.
 JPC: Josephson parametric converter (depicted as an on--off switch), with three ports---signal (\textsf{S}), idler (\textsf{I}), and pump.
 cQED: a transmon qubit (purple) dispersively coupled to a readout cavity (blue).
 SPA: SNAIL parametric amplifier \cite{Frattini2017,Frattini2018,Sivak2019}, which enables high-fidelity single-shot qubit readout.
 }
 \label{fig1}
\end{figure}

Although employing a different physical system, our experiment pays tribute to a classic measurement setup in radio astronomy invented by R.~H.~Dicke and R.~Beringer \cite{Dicke1946c}, whose block diagram is sketched in Fig.~\ref{fig1}(a).
Enclosed in blue is a microwave radiometer of the Dicke type \cite{Dicke1946a}, characterized by a two-way switch at the receiver input selecting between a black-body radiator at a reference temperature $T_\mathrm{ref}$, and a receiving antenna whose brightness is a mixture of extraterrestrial radiation and atmospheric background:
$T_\mathrm{ant} = \gamma  T_\mathrm{atm} + (1 - \gamma) T_\mathrm{ext}$,
with $\gamma$ denoting the atmospheric \emph{fractional absorption} \cite{Dicke1946b}.
After calibrating $T_\mathrm{ant}$ using the reference radiator and separately measuring $T_\mathrm{atm}$ and $\gamma$, one can extract $T_\mathrm{ext}$---the microwave temperature of the astronomical object.

Our measurement setup is depicted in Fig.~\ref{fig1}(b), with experimental parameters listed in Table~S1 of Ref.~\cite{suppl}.
At the system level, it can be viewed as a quantum-circuit realization of Dicke and Beringer's radio telescope.
The analogy is highlighted by the color code.
Our ``antenna,'' with a tunable resonant frequency $f_\mathrm{a}$ and a mode temperature $T_\mathrm{a}$, is a superconducting $LC$ resonator \cite{Xu2019a} anchored to a variable-temperature stage (VTS), whose temperature $T_\mathrm{VTS}$ can be adjusted between 1--2~K without affecting other parts of the dilution refrigerator.
The antenna mode is coupled with rate $\kappa_\mathrm{a,i}$ to the internal dissipative bath at $T_\mathrm{VTS}$, and with rate $\kappa_\mathrm{a,c}$ to a transmission line at $T_\mathrm{ext}\sim 0.1\ \si{K}$ \cite{VTS1}.
Moreover, an external broadband noise generator is able to raise the temperature of the transmission line by 0--2 K.
In the quantum regime, i.e., $k_\mathrm{B} T_j \lesssim hf_\mathrm{a}$ with $j$ denoting a mode or bath index, it is convenient to work with thermal photon populations in lieu of temperatures, which are linked by $\bar{n}_j = (e^{hf_\mathrm{a}/k_\mathrm{B} T_j}-1)^{-1}$.
The thermal population of the antenna mode is then given by
\begin{equation} \label{eq_cool}
\bar{n}_\mathrm{a} = \gamma \bar{n}_\mathrm{VTS} + (1-\gamma) (\bar{n}_\mathrm{ext} + \bar{n}_\mathrm{add})
\end{equation}
Here $\gamma \coloneqq \kappa_\mathrm{a,i} / (\kappa_\mathrm{a,i} + \kappa_\mathrm{a,c}) = \kappa_\mathrm{a,i} / \kappa_\mathrm{a} \sim 0.3$ is analogous to the ``fractional absorption,'' which is measured by a network analyzer through the ports \textsf{Probe~in} and \textsf{Probe~out}.
During the experiment, $T_\mathrm{VTS}$ is continuously monitored by a calibrated thermometer.
Therefore, the value of $\bar{n}_\mathrm{a}$ can be computed once $T_\mathrm{ext}$ is obtained.

The detector of our quantum radiometer is a standard cQED module in the strong dispersive regime, where photon-number fluctuations in a 3D readout cavity \cite{Paik2011} induce extra dephasing in a superconducting transmon qubit \cite{Cottet2002, Koch2007, Schreier2008}.
The intensity of the cavity field can then be deduced given its photon statistics.
For instance, with broadband thermal noise at the cavity input, this photon-induced-dephasing rate is given by \cite{Clerk2007}
\begin{equation} \label{eq_white}
\Gamma_\mathrm{th}(\bar{n}_\mathrm{r}^\mathrm{th}) = \frac{\kappa_\mathrm{r}}{2} \!\left[ \mathrm{Re}\sqrt{\left(1+\frac{i \chi}{\kappa_\mathrm{r}} \right)^2 + \frac{4i \chi \bar{n}_\mathrm{r}^\mathrm{th}}{\kappa_\mathrm{r} }  } -1 \right],
\end{equation}
where $\bar{n}_\mathrm{r}^\mathrm{th}$ is the cavity thermal population; $\kappa_\mathrm{r} = \kappa_\mathrm{r,c} + \kappa_\mathrm{r,i} $ is the cavity linewidth; $\chi/2\pi$ is the dispersive shift of the qubit frequency per cavity photon.
This dephasing rate can be measured via the Ramsey-fringe technique.
Dispersive qubit readout \cite{Narla2017} can be performed in reflection through the ports \textsf{cQED~in} and \textsf{cQED~out}.

The cQED detector is only sensitive to incoming photons in the vicinity of the qubit-readout frequency $f_\mathrm{r}$. To match it to the antenna resonator, a Josephson parametric converter (JPC) \cite{Bergeal2010a,Bergeal2010b,Roch2012} is pumped at the frequency difference of its \textsf{S} and \textsf{I} ports and operated as a lossless microwave frequency converter \cite{Abdo2013} with a bandwidth $\sim 100\ \si{\MHz}$.
Its conversion efficiency can be continuously adjusted from zero to unity.
Without the pump, the JPC provides over 40-dB isolation between its \textsf{S} and \textsf{I} ports, protecting the qubit from antenna radiation during its initialization, manipulation, and readout.

Our quantum radiometer shows clear advantages over its classical counterpart in sensing microwave radiation in the quantum regime.
First, by mapping the cavity population onto extra qubit dephasing, the cQED module directly measures the antenna radiation at the quantum level without passing it through any amplifier chain or demodulation circuit and without any sensitivity to vacuum noise.
The system noise is thus significantly suppressed.
Second, as qubit dephasing is caused by photon-number fluctuations in the readout cavity, our detector essentially probes the excitations of an electromagnetic field instead of its quadrature amplitudes, and is thus immune to the one-photon vacuum fluctuation, which is the inherent noise limit of classical radiometers with phase-preserving linear amplifiers \cite{Tucker1985,Bolgiano1961,Heffner1962,Caves1982}.
These two factors contribute to the precision of the following experimental results.

\section{Radiometry protocol}

\begin{figure} [b]
 \includegraphics[angle = 0, width = \columnwidth]{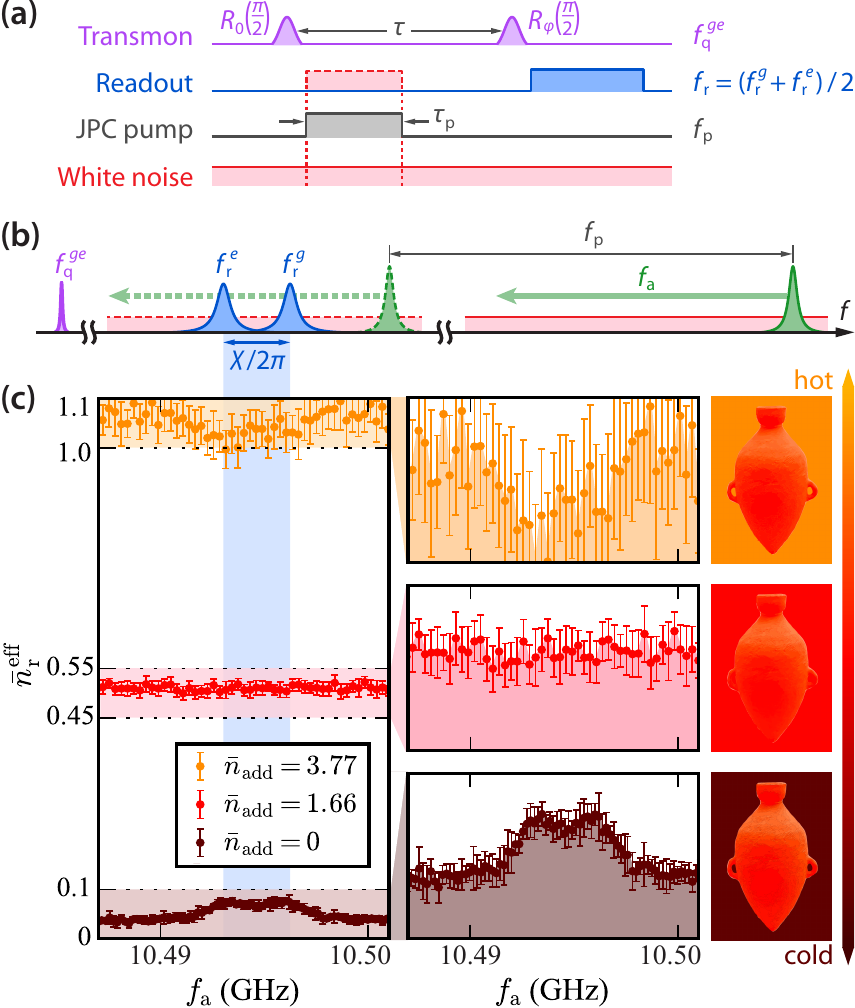}
 \caption{\textbf{(a)} Pulse sequences for noise-induced-dephasing measurements.
 For all data sets reported in this article, qubit rotations (purple) are completed in 200 ns; the second $\pi/2$ pulse is applied after the JPC pump (gray) by $\tau-\tau_\mathrm{p} = 1\ \si{\us}$; the single-shot qubit readout (blue) lasts for 1.08 \si{\us}.
 \textbf{(b)} Frequency landscape of the experiment. The solid green peak (right) belongs to the antenna resonator.
 The light green arrow indicates its frequency tunability.
 Along the frequency sweep, its down-converted image (dashed green peak) at $f_\mathrm{a} - f_\mathrm{p}$ at crosses the two resonances of the qubit-readout cavity (blue, $f_\mathrm{r}^{g(e)}=f_\mathrm{r}\pm \chi/4\pi$) separated by $\chi/2\pi$.
 The red floor refers to the added white noise.
 \textbf{(c)} Qubit-dephasing spectra at different added white-noise powers. The blue ribbon stands for the range $f_\mathrm{r}^e<f_\mathrm{a} - f_\mathrm{p}<f_\mathrm{r}^g$.
 The three zoomed-in plots, from top to bottom, show the radiative heating, equilibrium, and cooling regimes, respectively. The accompanying pictures illustrate these regimes through the color (temperature) contrasts between the amphorae and their backgrounds. For data in this subfigure, $\tau_\mathrm{p} = 0.54\ \si{\us}$ and $T_\mathrm{VTS}=1.03\ \si{\K}$.
 }
 \label{fig2}
\end{figure}

Our radiometry measurements are based on the Ramsey-interferometry experiment \cite{Ramsey1950,Vion2002}.
As shown in Fig.~\ref{fig2}(a), two $\pi/2$ qubit pulses are separated by a fixed delay $\tau$.
By varying their phase difference $\varphi$ and projectively measuring the qubit state afterwards, one can observe $2\pi$-periodic Ramsey fringes with an amplitude equal to $\exp [-\!\int_0^\tau \! \Gamma(t)\, \mathrm{d}t ]/2$, in which $\Gamma(t)$ is the instantaneous qubit decoherence rate \cite{suppl}.
When the JPC is in the isolation mode, $\Gamma(t)$ equals the inherent Ramsey decoherence rate $\Gamma_{2\mathrm{R}}=1/T_{2\mathrm{R}}=(4\pm0.3)\times 10^4\ \si{s}^{-1}$.
Contrarily, when the JPC is in the full-conversion mode, antenna radiation enters the cQED circuit while the residual thermal photons at port \textsf{I} responsible for $\Gamma_{2\mathrm{R}}$ are redirected to the port \textsf{S}.
(Here we assume that $\Gamma_{2\mathrm{R}}$ is solely caused by the residual photons coming from the input line.)
The ratio of the Ramsey amplitudes in these two circumstances equals $\exp [\int_0^\tau \! \Gamma_\mathrm{a}(t) \, \mathrm{d}t -  \Gamma_\mathrm{2R} \tau_\mathrm{p}]$, in which $\Gamma_\mathrm{a}(t)$ is the net dephasing rate induced by the antenna radiation, and $\tau_\mathrm{p}$ is the duration of the JPC-conversion pump, which also limits the frequency resolution of the radiometry protocol to $\sim \tau_\mathrm{p}^{-1}$.
We can thus experimentally obtain an average dephasing rate $\bar{\Gamma}_\mathrm{a}=\tau_\mathrm{p}^{-1} \int_0^\tau \! \Gamma_\mathrm{a}(t)\, \mathrm{d}t$, in which the integral limit and the prefactor are chosen such that $\bar{\Gamma}_\mathrm{a} = \Gamma_\mathrm{th}(\bar{n}_\mathrm{r}^\mathrm{th})$ if the antenna radiation is steady-state white noise with an average thermal population\ $\bar{n}_\mathrm{r}^\mathrm{th}$.
Finally, using Eq.~(\ref{eq_white}), we link $\bar{\Gamma}_\mathrm{a}$ to an \emph{effective} cavity thermal population $\bar{n}_\mathrm{r}^\mathrm{eff}$ as if the dephasing is induced by steady-state white noise.
Note that instead of measuring the frequency shifts of the zero-crossing points of Ramsey fringes, we essentially read out the fringe peaks where shot noise vanishes in the absence of thermal photons or intrinsic qubit decoherence.

In practice, we choose $\tau_\mathrm{p} \lesssim \mathrm{min}(T_\mathrm{2R}, T_1)/10$ to both ensure sufficient resolution on $\bar{n}_\mathrm{r}^\mathrm{eff}$ and mitigate qubit-relaxation events during the detection window.
The statistical error bars of $\bar{\Gamma}_\mathrm{a}$ and $\bar{n}_\mathrm{r}^\mathrm{eff}$ are contributed by both measurement imprecisions and the qubit $T_1$ fluctuations over time.
The former scale as $1/\sqrt{N_\mathrm{rep}}$, where $N_\mathrm{rep}$ is the number of repetition of the Ramsey sequence.
In Ref.~\cite{suppl}, we estimate the dynamic range of this radiometer to be $\sim 50\ \si{dB}$.

\section{Radiative cooling}

Following this protocol, we measured $\bar{n}_\mathrm{r}^\mathrm{eff}$ while sweeping the frequency of the antenna resonator.
Data are plotted in Fig.~\ref{fig2}(c).
When the down-converted antenna image is detuned from $f_\mathrm{r}^g$ and $f_\mathrm{r}^e$, the qubit senses the white noise from the \emph{external} bath with the thermal population $\bar{n}_\mathrm{ext} + \bar{n}_\mathrm{add}$ reflected by the antenna.
When the antenna image is aligned with $f_\mathrm{r}^g$ or $f_\mathrm{r}^e$, the qubit also senses the photons transmitted from the \emph{internal} bath at $T_\mathrm{VTS}=1.03\ \si{\K}$.
As photons from a hotter bath lead to more qubit dephasing, three distinct regimes were observed:
(i) If $T_\mathrm{add}=0$, the dephasing increases when the resonances are aligned, indicating $T_\mathrm{ext}<T_\mathrm{VTS}$.
(ii) If $T_\mathrm{add} \approx 1\ \si{\K}$, the dephasing is independent of $f_\mathrm{a}$, corresponding to $\bar{n}_\mathrm{ext} + \bar{n}_\mathrm{add} = \bar{n}_\mathrm{VTS}$.
(iii) At any higher $T_\mathrm{add}$, the dephasing decreases when the resonances are aligned, suggesting $\bar{n}_\mathrm{ext} + \bar{n}_\mathrm{add} > \bar{n}_\mathrm{VTS}$.
(i) and (iii) represent the radiative cooling and heating of the antenna mode by its external bath, respectively.

Our method for comparing $T_\mathrm{ext}$ and $T_\mathrm{VTS}$ by measuring thermal-photon-induced qubit dephasing is analogous to the disappearing-filament technique in pyrometry \cite{Holborn1903,Michalski2001}, which maps temperatures onto visible colors.
Illustrated by the pictures on the right of Fig.~\ref{fig2}(c), the peaked, flat, and dipped dephasing spectra ($\bar{n}_\mathrm{r}^\mathrm{eff}$--$f_\mathrm{a}$ curves) are associated with the amphorae \cite{amphora} being brighter, indistinguishable, and darker compared to the background radiation, respectively.
Temperature differences are manifested by the data contrasts in both measurements, although one is performed in a kiln and the other in a dilution fridge.

\begin{figure} [t]
 \includegraphics[angle = 0, width = \columnwidth]{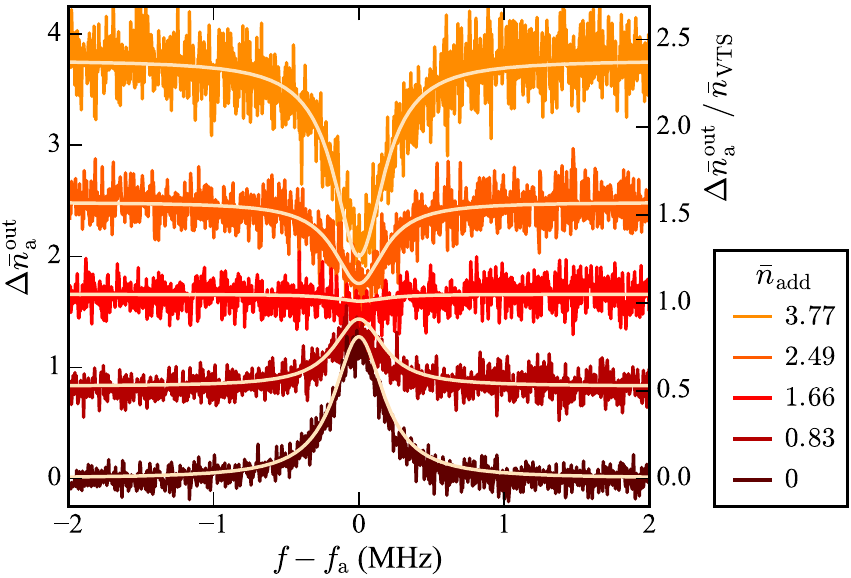}
 \caption{Radiative cooling and heating verified through classical radiometry.
 The theoretical curves (light yellow) are plotted based on the experimental values: $f_\mathrm{a} = 10.519\ \si{\GHz}$, $\kappa_\mathrm{a,c}/2\pi=0.30\ \si{\MHz}$, $\kappa_\mathrm{a,i}/2\pi=0.12\ \si{\MHz}$, and $T_\mathrm{VTS}=1.03\ \si{\K}$.
 }
 \label{fig4}
\end{figure}

To cross-check the above results, we performed a separate classical radiometry experiment by probing the power at \textsf{Probe out} [see Fig.~\ref{fig1}(b)] with a commercial spectrum analyzer \cite{rad_cool}.
In this experiment, the JPC is pumped at the frequency sum of its \textsf{S} and \textsf{I} ports and operated as a 23-dB quantum-limited phase-preserving amplifier.
The receiver chain is further equipped with a 33-dB HEMT amplifier on the 3.5-K stage of the dilution fridge and two 28-dB RF amplifiers at room temperature.
We first measured the background spectra within a 4-MHz detection window, and then tuned the antenna to the center of the window, took the spectra, subtracted the backgrounds, and referred the powers to the output photon fluxes per unit bandwidth of the antenna resonator \cite{suppl}.
Results are plotted in Fig.~\ref{fig4}, together with the theoretical curves $\Delta \bar{n}_\mathrm{a}^\mathrm{out} [f\!-\!f_\mathrm{a}] = \bar{n}_\mathrm{VTS} t_\mathrm{a}[f\!-\!f_\mathrm{a}] + \bar{n}_\mathrm{add}(1 - t_\mathrm{a}[f\!-\!f_\mathrm{a}] )$, in which $t_\mathrm{a}[f\!-\!f_\mathrm{a}]\! =\! 4 \gamma (1\!-\!\gamma) /\! \left[ 1 \!+\! 16\pi^2(f\!-\!f_\mathrm{a})^2 \!/ \!\kappa_\mathrm{a}^2 \right]$ denotes the Lorentzian transmission function of the resonator.
The topmost, middle, and bottommost traces were measured with the same added white-noise powers as the three qubit-dephasing spectra in Fig.~\ref{fig2}(c).
The Lorentzian peak evolves into a dip as $\bar{n}_\mathrm{add}$ increases, which manifests the competition between the internal and external baths, and distinguishes the regimes of radiative cooling and heating.
The agreement between the data and the theory that assumes a constant $T_\mathrm{VTS}=1.03\ \si{K}$ confirms the internal bath was not influenced while the external bath was being heated by the white-noise source.
Therefore, readouts of the VTS thermometer can be reliably used in the calibration procedure of the radiometer.

\section{Noise analysis}

\begin{figure} [b]
 \includegraphics[angle = 0, width = \columnwidth]{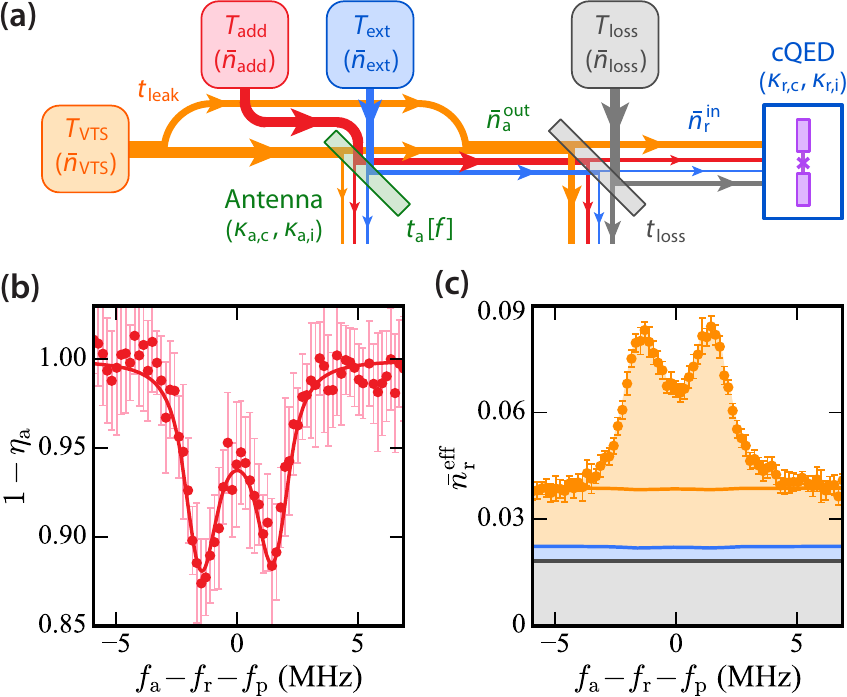}
 \caption{
  \textbf{(a)} Noise model of the radiometer.
 Left beam splitter (green): tunable antenna resonator.
 Right beam splitter (gray): ohmic loss along the coaxial cables and circulators between the resonator and the cQED module.
 Four colored boxes represent the thermal baths.
 Yellow: internal bath in equilibrium with the VTS.
 Red: white-noise source.
 Blue: external bath in the transmission line.
 Gray: dissipation between the antenna and the cQED cavity.
 \textbf{(b)} Normalized slopes and \textbf{(c)} vertical intercepts of the $\bar{n}_\mathrm{r}^\mathrm{eff}$--$\bar{n}_\mathrm{add}$ lines
 at different antenna frequencies with $T_\mathrm{VTS}=1.03\ \si{\K}$.
 {
 The solid red curve in \textbf{(b)} is the theoretical prediction based on the experimental values of $\kappa_\mathrm{a,c}$, $\kappa_\mathrm{a,i}$, $\kappa_\mathrm{r}$, and $\chi$, with no fitting parameter.
 }
 In \textbf{(c)}, contributions of different baths are indicated using the same color code as that in \textbf{(a)}.
 The orange line separates the VTS leakage (below) and the transmitted radiation from the resonator (above).
 For all data in this figure, $\tau_\mathrm{p}=1.08\ \si{\us}$.
 }
 \label{fig3}
\end{figure}

We model our radiometer using the signal-flow diagram in Fig.~\ref{fig3}(a) when the JPC is in the full-conversion mode, and write the photon flux per unit bandwidth at the input of the qubit-readout cavity as
\begin{equation} \label{eq_nin}
\bar{n}_\mathrm{r}^\mathrm{in}[f\!-\!f_\mathrm{a}] =
\bar{n}_\mathrm{a}^\mathrm{out} [f\!-\!f_\mathrm{a}] t_\mathrm{loss} + \bar{n}_\mathrm{loss} (1 - t_\mathrm{loss}),
\end{equation}
where
$\bar{n}_\mathrm{a}^\mathrm{out} [f\!-\!f_\mathrm{a}] = \bar{n}_\mathrm{VTS} (t_\mathrm{a}[f\!-\!f_\mathrm{a}] + t_\mathrm{leak} ) + (\bar{n}_\mathrm{ext} + \bar{n}_\mathrm{add} ) (1 - t_\mathrm{a}[f\!-\!f_\mathrm{a}])$ refers to the output photon flux per unit bandwidth of the antenna resonator.
In Eq.~(\ref{eq_nin}), the term proportional to $\bar{n}_\mathrm{loss}$ represents the noise added by the cables, connectors, and circulators between the resonator and the cQED module;
the term containing $\bar{n}_\mathrm{VTS} t_\mathrm{leak}$ represents the thermal leakage from the VTS.
Together with the shot noise associated with the inherent qubit decoherence, they constitute the \textit{system noise} of the radiometer---the total noise at the output of the radiometer in the absence of input signals.
Other terms, proportional to either $t_\mathrm{a}$ or $1-t_\mathrm{a}$, are associated with photons from the internal or external bath of the resonator.

Unlike a standard square-law detector that measures the integral of impinging photon flux within its detection bandwidth, our cQED module does not directly detect $\bar{n}_\mathrm{r}^\mathrm{in} [f_\mathrm{r}^{g(e)} \!+\! f_\mathrm{p} \!-\! f_\mathrm{a}]$.
Instead, as we derive in Ref.~\cite{suppl}, the measured $\bar{n}_\mathrm{r}^\mathrm{eff}$ in our experiment can be expressed as
\begin{equation} \label{eq_color}
\begin{split}
\bar{n}_\mathrm{r}^\mathrm{eff}  \! &\left[f_\mathrm{r} \!+\! f_\mathrm{p} \!-\! f_\mathrm{a}, \tau_\mathrm{p} \right] \!=\! \left\{ \bar{n}_\mathrm{VTS}\, t_\mathrm{loss}\, \eta_\mathrm{a}\! \left[f_\mathrm{r} \!+\! f_\mathrm{p} \!-\! f_\mathrm{a}, \tau_\mathrm{p} \right] \right. \\
& + \left(\bar{n}_\mathrm{ext} + \bar{n}_\mathrm{add} \right) t_\mathrm{loss}\! \left(1-\eta_\mathrm{a} \! \left[f_\mathrm{r} \!+\! f_\mathrm{p} \!-\! f_\mathrm{a}, \tau_\mathrm{p} \right]\right)\\
& + \left. \bar{n}_\mathrm{loss} (1-t_\mathrm{loss}) + \bar{n}_\mathrm{VTS} \,t_\mathrm{leak}\, t_\mathrm{loss} \right\} \kappa_\mathrm{r}^{-2} \kappa_\mathrm{r,c} \kappa_\mathrm{a},
\end{split}
\end{equation}
in which the dimensionless function $\eta_\mathrm{a}$ summarizes the response of the cQED detector to the pulsed noise with a Lorentzian spectrum.
In Fig.~\ref{fig3}(b), we plot the experimental data of $\eta_\mathrm{a}$ together with our theoretical predictions \cite{suppl}.
By measuring $\bar{n}_\mathrm{r}^\mathrm{eff}$ with different values of $\bar{n}_\mathrm{VTS}$ and $\bar{n}_\mathrm{add}$, we calibrated the system parameters, obtained $\bar{n}_\mathrm{ext}$, and computed $\bar{n}_\mathrm{a}$ using Eq.~(\ref{eq_cool}).
The results are listed in Table~\ref{tab2}, where the system noise is given by $\bar{n}_\mathrm{sys} = \bar{n}_\mathrm{para} + \bar{n}_\mathrm{shot}$, in which $\bar{n}_\mathrm{para} \coloneqq \bar{n}_\mathrm{VTS} t_\mathrm{leak} + \bar{n}_\mathrm{loss} (1-t_\mathrm{loss})/ t_\mathrm{loss}$ denotes the parasitic white noise referred to the antenna input, and $\bar{n}_\mathrm{shot}$, derived in Ref.~\cite{suppl}, is the shot noise due to the inherent qubit decoherence.
We obtain $T_\mathrm{sys}=0.31\ \si{\K} \sim hf/2k_\mathrm{B}$, which is half of the minimum noise temperature of an ideal quantum-limited phase-preserving linear amplifier.
Radiative cooling is manifested again through $\bar{n}_\mathrm{ext} \sim \bar{n}_\mathrm{VTS}/100$ and $\bar{n}_\mathrm{a}\sim\bar{n}_\mathrm{VTS}/3$.
As illustrated in Fig.~\ref{fig3}(c), qubit dephasing is contributed by the white system noise (gray and lower orange), and the frequency-dependent radiation from the internal (upper orange) and external (blue) baths of the resonator. The shares of the latter two mechanisms in $\bar{n}_\mathrm{r}^\mathrm{eff}$ are proportional to $\eta_\mathrm{a}$ and $1-\eta_\mathrm{a}$, respectively.

As a comparison, the classical radiometry experiment described in the previous section using a quantum-limited parametric amplifier yields $\bar{n}_\mathrm{ext}^\mathrm{cl} = 0.02^{+0.06}_{-0.02}$ and $T_\mathrm{sys}^\mathrm{cl}=(1.02\pm0.01)\ \si{K} \sim 2hf/k_\mathrm{B}$ \cite{rad_cool}.
The advantages of the qubit radiometer in precision and system noise are thus clearly demonstrated.
Moreover, as shown in Ref.~\cite{suppl}, viewed from the point of photon counting, the performance of our qubit radiometer is comparable to state-of-the-art microwave single-photon detectors in the limit of small input-photon numbers.

\begin{table} [t]
\centering
\begin{tabular}{>{\centering\arraybackslash}m{1.8cm}
>{\centering\arraybackslash}m{2.25cm}|
>{\centering\arraybackslash}m{2.25cm}
>{\centering\arraybackslash}m{1.8cm}}
 \hline
 \hline
 $\bar{n}_\mathrm{loss}$ & $T_\mathrm{loss}$ (K) & $\bar{n}_\mathrm{ext}$ & $T_\mathrm{ext}$ (K) \\
 \hline
 $0.09\pm0.01$ & $0.20 \pm 0.01$ & $0.014\pm 0.005$ & $0.11\pm0.01$  \\
 \hline
 \hline
 $\bar{n}_\mathrm{para}$ & $T_\mathrm{para}$ (K) & $\bar{n}_\mathrm{a}$ & $T_\mathrm{a}$ (K)\\
 \hline
 $0.16 \pm 0.02$ & $0.25 \pm 0.01$ & $0.49\pm0.01$ & $0.45 \pm 0.01$ \\
 \hline
  \hline
 $\bar{n}_\mathrm{sys}$ & $T_\mathrm{sys}$ (K) & $t_\mathrm{leak}$ & $t_\mathrm{loss}$\\
  \hline
 $0.25 \pm 0.02$ & $0.31 \pm 0.01$ & $0.046 \pm 0.005$ & $0.52\pm0.06$ \\
 \hline
 \hline
\end{tabular}
\caption{Radiometric results. The values of $\bar{n}_\mathrm{para}$ ($T_\mathrm{para}$), $\bar{n}_\mathrm{sys}$ ($T_\mathrm{sys}$), and $\bar{n}_\mathrm{a}$ ($T_\mathrm{a}$) are based on $T_\mathrm{VTS}=1.03\ \si{K}$ ($\bar{n}_\mathrm{VTS}=1.59$).
Thermal populations and temperatures are linked here by $\bar{n}_j = (e^{hf_\mathrm{a}/k_\mathrm{B} T_j}-1)^{-1}$.
}
\label{tab2}
\end{table}

\section{Perspectives}

Our radiometer comprises a tunable antenna resonator and a fixed-frequency narrowband detector (cQED module) matched by a broadband frequency converter (JPC).
Although this article reports the experiment of probing two white thermal baths at $T_\mathrm{ext}$ and $T_\mathrm{VTS}$, in which the JPC-conversion frequency was fixed while the antenna frequency was varied,
one can also sweep the conversion frequency together with the antenna frequency and thus use the qubit to sense a colored bath or radiator.
The JPC-conversion curve then needs to be calibrated at every pump frequency.
In this ``sweeping mode,'' the frequency range of this radiometer is broadened from $\kappa_\mathrm{r}\sim 1\ \si{MHz}$ to the JPC-conversion bandwidth $\sim 100 \ \si{MHz}$.
Gigahertz bandwidth would be achievable by if the JPC is replaced with a broadband three-wave-mixing parametric device \cite{Zorin2016}.

Two applications are within sight:
First, our setup can be used to study the thermodynamics of hybrid quantum systems, such as verifying the ground-state cooling of an electro-optomechanical photon converter \cite{Zhong2019} or a spin ensemble \cite{spin_cool}.
Second, superconducting qubits have been envisaged to replace low-noise amplifiers as itinerant photon detectors in microwave-cavity-based axion experiments \cite{Lamoreaux2013,Zheng2016,Dixit2018}.
Considering the resemblance of our measurement setup to an axion detector \cite{Asztalos2010,Du2018,Brubaker2017}, this radiometry protocol is useful for detecting this hypothetical particle and constraining theories of dark matter.

\section{Acknowledgements}
We acknowledge V. V. Sivak for fabricating the SNAIL parametric amplifier, and P. Bertet, A. A. Clerk, L. DiCarlo, Liang Jiang, S. K. Lamoreaux, and K. W. Lehnert for insightful discussions.
Z.W. thanks P. Reinhold and {S. O. Mundhada} for their assistance on FPGA electronics.
Use of facilities was supported by the Yale Institute for Nanoscience and Quantum Engineering and the Yale School of Engineering and Applied Science clean room.
This research was supported by the US Army Research Office (Grants No.~W911NF-18-1-0020 and W911NF-18-1-0212), the US Air Force Office of Scientific Research (Grant No.~FA9550-15-1-0029), and the National Science Foundation (Grant No.~DMR-1609326).

\bibliography{radiometry_bib}

\begin{thebibliography}{69}
\bibitem [{sen()}]{sensitivity}%
  \BibitemOpen
  \href@noop {} {}\bibinfo {note} {The sensitivity of a total-power radiometer
  is written as $\delta T = T_\mathrm{sys}\sqrt{(B \tau_\mathrm{int})^{-1} +
  (\delta G/G)^2}$, in which $\tau_\mathrm{int}$ is the integration time; $B$
  is the detection bandwidth; $G$ and $\delta G$ are the receiver gain and its
  fluctuation.}\BibitemShut {Stop}%
\end{thebibliography}

\end{document}


\title{Supplemental Material for \\``Quantum Microwave Radiometry with a Superconducting Qubit''}











\maketitle

\beginsup

\section{Experimental Details}

\begin{figure*}[t]
\includegraphics[width=\textwidth]{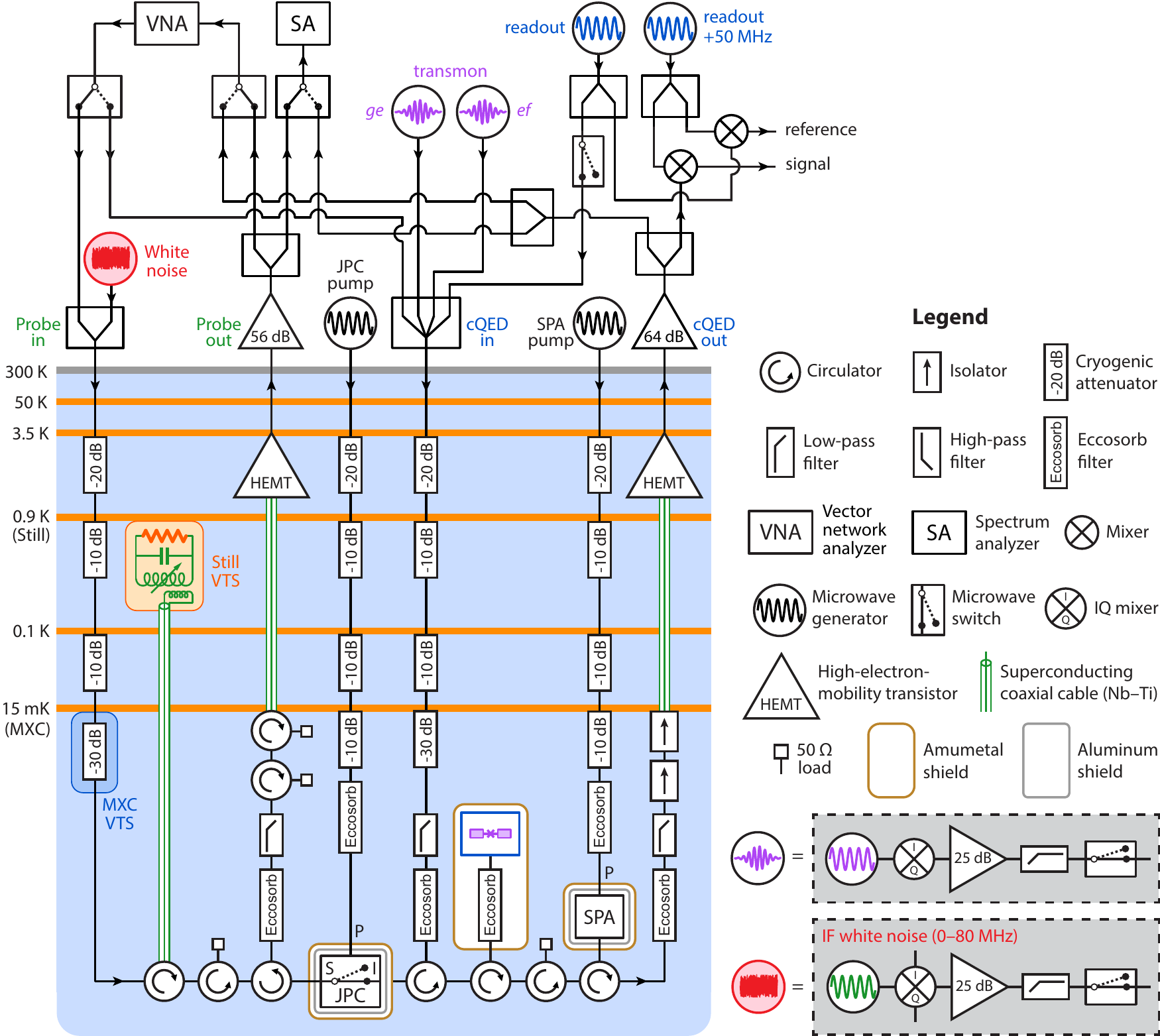}
\caption{Detailed cryogenic wiring diagram with simplified room-temperature electronics.
Envelopes of qubit pulses are generated by the 500 Msample/s analog outputs of a field programmable gate array (FPGA) card and sent to the IQ mixers.
Qubit-readout pulses are shaped by a high-frequency microwave switch.
The 50-MHz \textsf{signal} and \textsf{reference} outputs of the heterodyne interferometer are digitized and processed by the same FPGA card.
}
\label{fig_fridge}
\end{figure*}

A detailed drawing of our experimental setup is shown in Fig.~\ref{fig_fridge}.
Low-temperature measurements were performed in a cryogen-free dilution refrigerator.
Our transmon qubit consists of an Al/AlO$_x$/Al Josephson junction connected to a pair of rectangular aluminum pads.
The junction is defined on a sapphire chip using the bridge-free electron-beam lithography technique \cite{Rigetti2009, Lecocq2011} and fabricated via double-angle evaporation.
The readout resonator is a 3D indium-plated copper cavity, which has a superconducting surface and a high-thermal-conductivity bulk at low temperatures.
This cQED module is housed in a mu-metal (Amumetal 4K) magnetic shield and thermalized to the 15-mK mixing-chamber (MXC) stage.
Homemade Eccosorb filters are installed on the input and output lines, and, in addition, on the cavity-to-SMA coupler inside the mu-metal shield to block spurious high-frequency radiation and improve qubit coherence times \cite{Serniak2019}.
Using standard pulse sequences, we measured $T_1$, Ramsey $T_{2\mathrm{R}}$, and Hahn echo $T_{2\mathrm{e}}$, and computed the inherent qubit dephasing rate $\Gamma_\phi = T_\mathrm{2R}^{-1} - (2T_1)^{-1} = (3\pm 0.3) \times 10^{4}\ \mathrm{s}^{-1}$.
The fact $T_{2\mathrm{R}} \approx T_{2\mathrm{e}}$ indicates that there is little low-frequency noise and thus that qubit dephasing is primarily contributed by residual thermal photons in the readout cavity.
Attributing $\Gamma_\phi$ to the residual thermal population in the fundamental mode of the readout cavity, we obtain $\bar{n}_\mathrm{r}^\mathrm{th}=(7\pm0.7)\times 10^{-3}$.
Based on the results of our previous experiment \cite{Wang2019}, we assume all these parasitic thermal photons solely come from the input line of the cQED module.

The Josephson parametric converter (JPC) and the SNAIL parametric amplifier (SPA) are enclosed in aluminum cans inside their own mu-metal shields, and are both thermalized to the MXC stage.
Magnetic fluxes are applied to the Josephson-junction loops through DC coils mounted inside the aluminum cans.
The bias fluxes are chosen such that the JPC idler resonator and the SPA resonator are both aligned to the qubit-readout cavity.
Consequently, the JPC converts antenna radiation to the vicinity of the qubit-readout frequency $f_\mathrm{r}=(f_\mathrm{r}^g+f_\mathrm{r}^e)/2$ with a close-to-unity efficiency (see Sec.~\ref{sec_JPC}).
By pumping the SPA at $2f_\mathrm{r}+10\ \si{\MHz}$, we operate it as a quantum-limited degenerate phase-preserving amplifier with 14-dB gain for qubit-readout signals.
The fidelity of the single-shot dispersive readout is optimized to be above 0.95, which is mainly limited by the less-than-expected SPA gain and can be improved by a higher-gain device \cite{Frattini2018,Sivak2019}, with which a readout fidelity of 0.975 has been previously recorded in the same measurement setup.

Our tunable antenna resonator is anchored to a copper stage (variable-temperature stage, VTS), which is itself weakly thermalized to the still stage of the dilution fridge via a section of copper wire.
The temperature of the VTS can be raised by a heater and monitored in time by a ruthenium-oxide thermometer.
When the heater is off, we measured $T_\mathrm{still}=0.87\ \si{K}$ and $T_\mathrm{VTS}=1.03\ \si{K}$.
When the heater current is changed, the VTS will reach its new stable temperature with a time constant $\sim 15\ \si{\min}$.
While varying $T_\mathrm{VTS}$ between 1.03 K and 2.2 K, we did not observe any change on the temperatures of other fridge stages beyond their normal fluctuations.

The last 30-dB attenuator on the antenna input line is anchored to a copper VTS that is constructed similarly to the one holding the antenna resonator but is thermalized to the MXC stage instead.
In the absence of input signals, the thermometer attached to the MXC-stage VTS reads $70\ \si{mK}$, which sets a lower bound of $T_\mathrm{ext}$.
During the experiment, the heater on the MXC-stage VTS remained off.
The temperature of the external bath of the antenna is thus only raised by the added noise with 80-MHz bandwidth near the antenna frequency.

\begin{table} [t]

\centering
\begin{tabular}{>{\centering\arraybackslash}m{3.5cm}|>{\centering\arraybackslash}m{2.65cm}
>{\centering\arraybackslash}m{2cm}
}
 \hline
 \hline
 \multirow{7}{*}{Transmon qubit} & $f_\mathrm{q}^{ge}$ (GHz) & 4.6820 \\
 & $f_\mathrm{q}^{ef}$ (GHz) & 4.4487 \\
 & $\chi/2\pi$ (MHz) & 3.1 \\
 & $T_1$ ($\si{\us}$) & $71\pm2$ \\
 & $T_\mathrm{2R}$ ($\si{\us}$) & $24\pm2$ \\
 & $T_\mathrm{2e}$ ($\si{\us}$) & $27\pm2$ \\
 & $P_e^\mathrm{ini}$ & $0.03\pm0.002$ \\
 \hline

 \multirow{4}{*}{cQED readout cavity} & $f_\mathrm{r}$ (GHz) & 7.6011 \\
 & $\kappa_\mathrm{r}/2\pi$ (MHz) & 0.83 \\
 & $\kappa_\mathrm{r,i}/2\pi$ (MHz) & 0.06 \\
 & $\kappa_\mathrm{r,c}/2\pi$ (MHz) & 0.77\\
 \hline

 \multirow{4}{*}{Antenna resonator} & $f_\mathrm{a}$ (GHz) & 10.48--10.52\\
 & $\kappa_\mathrm{a}/2\pi$ (MHz) & 0.37--0.42\\
 & $\kappa_\mathrm{a,i}/2\pi$ (MHz) & 0.10--0.14\\
 & $\kappa_\mathrm{a,c}/2\pi$ (MHz) & 0.26--0.30\\
 \hline

 \multirow{5}{*}{JPC} & $f_\mathrm{S}$ (GHz) & 10.52\\
 & $\kappa_\mathrm{S}/2\pi\ \!$(MHz) & 56\\
 & $f_\mathrm{I}$ (GHz) & 7.61\\
 & $\kappa_\mathrm{I}/2\pi\ \!$(MHz) & 83\\
 & $f_\mathrm{p}$ (GHz) & 2.8935\\
 \hline
 \hline
\end{tabular}

\caption{Summary of device parameters.
$f_\mathrm{q}^{ge}$ and $f_\mathrm{q}^{ef}$: $ge$- and $ef$-transition frequencies of the transmon qubit.
$P_e^\mathrm{ini}$: excited-state population of the transmon in the absence of drives.
Subscript ``i'': internal dissipation; ``c'': external coupling. Total linewidth of the readout cavity (antenna resonator): $\kappa_\mathrm{r(a)} = \kappa_\mathrm{r(a),i}+\kappa_\mathrm{r(a),c}$.
}
\label{tab1}
\end{table}

Key parameters of this experiment are listed in Table~\ref{tab1}.
Frequencies and linewidths of the JPC resonators were characterized with weak probe tones without pumps.

\section{Frequency-Tunable Antenna Resonator}

\begin{figure}[t]
\includegraphics[width=\columnwidth]{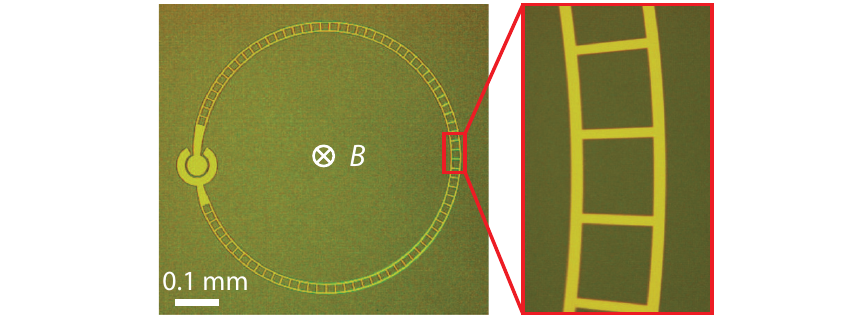}
\caption{Optical micrograph of the antenna resonator. The total inductance of the device comprises a geometric part and a kinetic part. The zoomed-in view shows the square holes on the inductor arm responsible for the frequency tunability of the antenna resonator.}
\label{fig_micro}
\end{figure}

\begin{figure}[t]
\includegraphics[width=\columnwidth]{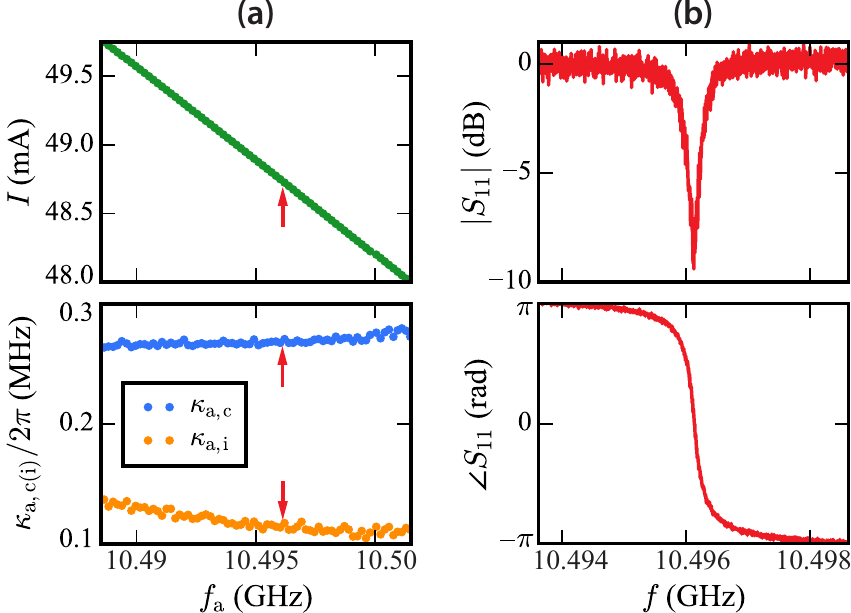}
\caption{(a) Antenna parameters measured during the frequency sweep of Figs.~4(b) and 4(c). Top: magnetic-coil current versus antenna frequency, where $I=60\ \si{mA}$ corresponds to $B\sim 1\ \si{mT}$ perpendicular to the device plane.
Bottom: $\kappa_\mathrm{a,c}$ and $\kappa_\mathrm{a,i}$ at different antenna frequencies.
(b) Antenna reflection at $I=48.765\ \si{mA}$, where $f_\mathrm{a}=10.4961\ \si{GHz}$, $\kappa_\mathrm{a,c}=0.27\ \si{MHz}$, $\kappa_\mathrm{a,i}=0.12\ \si{MHz}$, and $\gamma=0.31$. These values are pointed out in (a) by the red arrows.
}
\label{fig_ouro}
\end{figure}

The antenna of our radiometer is a 2D niobium-nitride superconducting $LC$ resonator deposited on a sapphire substrate, which is inductively coupled to a niobium--titanium superconducting transmission line.
As shown in Fig.~\ref{fig_micro}, square holes on the circular inductor arm allow the kinetic inductance of the resonator to be increased by an external magnetic field perpendicular to the device plane, down-shifting the resonator frequency \cite{Xu2019a}.
In this experiment, the antenna frequency can be lowered from 10.7 GHz to 10.4 GHz by a $\sim 1\ \si{mT}$ magnetic field.
In the upper panel of Fig.~\ref{fig_ouro}(a) is the magnetic-coil current plotted against the antenna frequency within the range of our radiometry measurements.
A frequency hysteresis $\sim 30\ \si{kHz}$ can be observed if one sweeps the magnetic field bi-directionally.
To address this hysteresis, each antenna-frequency point in the qubit-dephasing spectra ($\bar{n}_\mathrm{r}^\mathrm{eff}$--$f_\mathrm{a}$ curves) reported in the Main Text was measured by a network analyzer immediately before the Ramsey experiment.
As shown in Fig.~\ref{fig_ouro}(b), $\kappa_\mathrm{a,c}$ and $\kappa_\mathrm{a,i}$ weakly depend on the antenna frequency, giving $\gamma \coloneqq \kappa_\mathrm{a,i}/\kappa_\mathrm{a} \sim 0.3$.

\section{Frequency Conversion through a Josephson Parametric Converter} \label{sec_JPC}

The JPC is a frequency-tunable non-degenerate three-wave-mixing device for converting or amplifying microwave signals close to the quantum limit \cite{Flurin2014,Sliwa2016}. It consists of a Josephson ring modulator (JRM) embedded in two crossed $\lambda/2$ resonators denoted signal (\textsf{S}) and idler (\textsf{I}).
To use the JPC as a quantum-limited frequency converter, we apply a pump at the frequency difference of its \textsf{S} and \textsf{I} resonators: $f_\mathrm{p}=f_\mathrm{S} - f_\mathrm{I}$.
Due to the nonlinearity of the JRM, when the pump is applied, $f_\mathrm{S}$ and $f_\mathrm{I}$ are down-shifted by $\sim 10\ \si{MHz}$ from their linear-response values recorded in Table~\ref{tab1}.
The pump frequency should thus be optimized within the range of a few megahertz to yield the maximum conversion efficiency between the antenna and the cQED module.

\begin{figure}[t]
\includegraphics[width=\columnwidth]{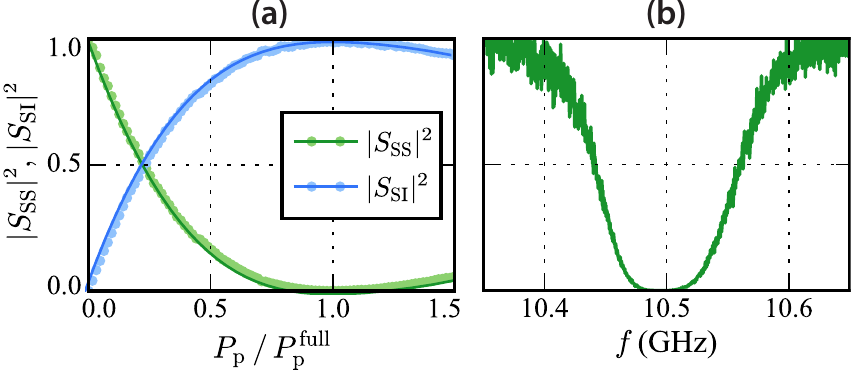}
\caption{(a) JPC reflection and conversion coefficients as functions of the pump power $P_\mathrm{p}$. Green (blue) curve: polynomial fit of the $|S_\mathrm{SS(SI)}|^2$ data. (b) \textsf{S}-port reflection spectrum at the full-conversion point ($P_\mathrm{p}=P_\mathrm{p}^\mathrm{full}$).}
\label{fig_jpc}
\end{figure}

The JPC-conversion curves are calibrated with $f_\mathrm{p}=2.8935\ \si{GHz}$.
Using the method of Ref.~\cite{Abdo2013}, we first applied to the port \textsf{Probe in} a continuous-wave signal at 10.4961 GHz and monitored the output signal at the same frequency with a spectrum analyzer at the port \textsf{Probe out}.
The green dots in Fig.~\ref{fig_jpc}(a) are the reflection coefficients at different pump powers, which are computed by normalizing the output powers at \textsf{Probe out}, with the power measured with $P_\mathrm{p}=0$ mapped to $|S_\mathrm{SS}|=1$ and the noise floor mapped to $|S_\mathrm{SS}|=0$.
As shown in Fig.~\ref{fig_jpc}(b), we measured the reflection spectrum with a network analyzer when the JPC is in the full-conversion mode, and verified $|S_\mathrm{SS}|^2<0.001$ within the antenna-frequency range of our experiment.
To calibrate the JPC-conversion curve (blue), we pumped the JPC at its half-conversion point ($|S_\mathrm{SS}|^2=0.5$), applied to the port \textsf{cQED in} a continuous-wave signal at 7.6026 GHz, and adjusted its power such that the output power at \textsf{Probe out} is the same given either input (\textsf{Probe in} or \textsf{cQED in}).
Then by normalizing the converted power at \textsf{Probe out} in the same manner as the reflected power, we obtained the conversion coefficients plotted as the blue dots in Fig.~\ref{fig_jpc}(a).
The reflection and conversion curves are symmetric about $|S|^2=0.5$.
At the full-conversion point ($P_\mathrm{p}=P_\mathrm{p}^\mathrm{full}$), the maximum conversion efficiency is close to unity ($|S_\mathrm{SI}|^2 > 0.99$).
We thus model the JPC between its isolation and full-conversion mode as a near-ideal microwave switch with a response time $\sim 1/\kappa_\mathrm{S(I)} < 3\ \si{ns}$.

\section{Noise-Induced-Dephasing Experiment} \label{sec_ramsey}

\subsection{Ramsey experiment}

\begin{figure}[t]
\includegraphics[width=\columnwidth]{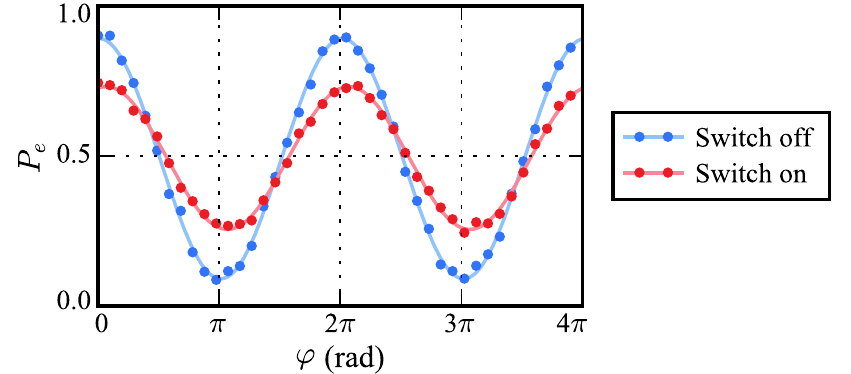}
\caption{Ramsey oscillations with and without antenna radiation. The JPC-conversion pump lasts for $\tau_\mathrm{p}=1.08\ \si{\us}$.}
\label{fig_ramsey}
\end{figure}

The noise-induced-dephasing measurement in this work is based on the Ramsey-interferometry experiment. Following the pulse sequences in Fig.~2(a), we show two sinusoidal Ramsey-oscillation curves in Fig.~\ref{fig_ramsey}, corresponding to the JPC switch being constantly off (blue) and turned on for duration $\tau_\mathrm{p}$ (red), respectively.
Their amplitudes
\begin{align}
A_\mathrm{off} = \frac{1}{2} e^{-\Gamma_\mathrm{2R} \tau}
\end{align}
and
\begin{align} \label{eq_Aon}
A_\mathrm{on} = \frac{1}{2} \exp \left\{- \int_0^\tau \!\! \left[ \Gamma_\mathrm{a}(t) + \Gamma_\mathrm{2R} \right] \mathrm{d}t + \Gamma_\mathrm{2R} \tau_\mathrm{p} \right\}
\end{align}
are related to the additional qubit dephasing induced by the antenna radiation via
\begin{equation} \label{eq_ramsey}
\frac{A_\mathrm{off}}{A_\mathrm{on}} = \exp \left[\int_0^\tau \! \Gamma_\mathrm{a}(t)\, \mathrm{d}t  - \Gamma_\mathrm{2R}\tau_\mathrm{p} \right] = e^{(\bar{\Gamma}_\mathrm{a} - \Gamma_\mathrm{2R}) \tau_\mathrm{p}},
\end{equation}
which is independent of the inherent qubit decoherence rate $\Gamma_\mathrm{2R}=1/T_\mathrm{2R}$.
In Eq.~(\ref{eq_Aon}), the term $\Gamma_\mathrm{2R}\tau_\mathrm{p}$ is added to the exponent to account for the bi-directional conversion of the JPC. Namely, while the JPC is in the full conversion mode, the residual thermal photons causing $\Gamma_{2\mathrm{R}}$, which are assumed to come from the qubit input line, are converted from port \textsf{I} to port \textsf{S}, and therefore have no contribution to qubit dephasing during $\tau_\mathrm{p}$.
We then obtain the effective thermal population of the qubit-readout cavity from the average antenna-induced-dephasing rate
\begin{equation} \label{eq_white}
\begin{split}
\bar{\Gamma}_\mathrm{a} = \frac{\kappa_\mathrm{r}}{2}\!\!\left[ \mathrm{Re} \sqrt{\!\left(\!1 \!+\!\frac{i \chi}{\kappa_\mathrm{r}}\! \right)^2 \!\!+\! \frac{4i \chi \bar{n}_\mathrm{r}^\mathrm{eff}}{\kappa_\mathrm{r}}} \!-\! 1 \right] \!\!
\xrightarrow{\bar{n}_\mathrm{r}^\mathrm{eff} \ll 1} \! \frac{\chi^2 \kappa_\mathrm{r} \bar{n}_\mathrm{r}^\mathrm{eff}}{\chi^2 + \kappa_\mathrm{r}^2 }.
\end{split}
\end{equation}
If the antenna noise is white, this definition of $\bar{n}_\mathrm{r}^\mathrm{eff}$ makes it equal to the actual thermal population of the qubit-readout mode.

\begin{figure}[b]
\includegraphics[width=\columnwidth]{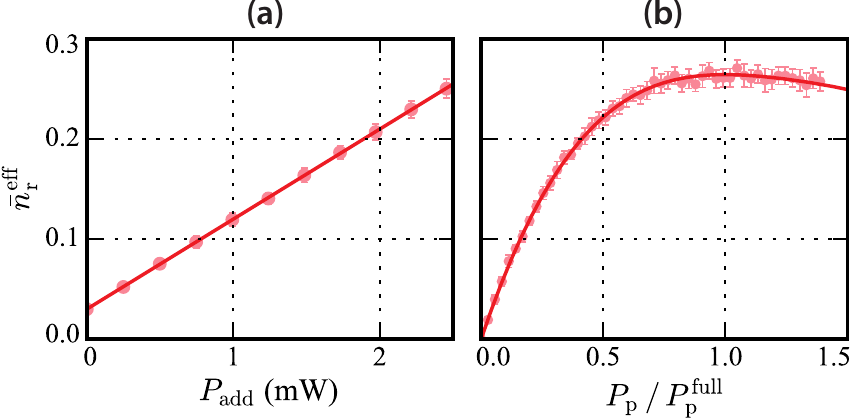}
\caption{White-noise-induced qubit dephasing.
(a) $\bar{n}_\mathrm{r}^\mathrm{eff}$ against white-noise-generator power at room temperature. The JPC is close to its full-conversion mode. (b) $\bar{n}_\mathrm{r}^\mathrm{eff}$ versus JPC pump power given $P_\mathrm{add}=2.5\ \si{mW}$.
The solid red curve is a rescaled JPC-conversion curve.
}
\label{fig_nid}
\end{figure}

We note that the qubit excited-state population and the infidelity of the single-shot qubit readout lower the Ramsey-oscillation contrast from unity even in the absence of dephasing channels.
But we are able to correct these offsets through dividing $A_\mathrm{on}$ by $A_\mathrm{off}$ instead of extracting $\bar{n}_\mathrm{r}^\mathrm{eff}$ directly from the Ramsey amplitudes.
Moreover, as can be found in Fig.~\ref{fig_ramsey}, the switch-on curve shows a phase shift compared to the switch-off curve.
This demonstrates the AC Stark shift of the qubit frequency induced by the cavity photons.
We can similarly extract $\bar{n}_\mathrm{r}^\mathrm{eff}$ from Stark-shift data.
The results match those obtained using dephasing rates but have larger error bars (data not shown).
We therefore used the dephasing data in the radiometry measurements of our experiment.

\subsection{Photon-number calibration}
We use Eq.~(\ref{eq_white}) to calibrate the temperature $T_\mathrm{add}$, or equivalently, the thermal population $\bar{n}_\mathrm{add}$ of the external white-noise source referred to the input of the antenna resonator.
To do this, we adjust the antenna frequency such that $|f_\mathrm{a} - f_\mathrm{r} - f_\mathrm{p}| \gg \chi/2\pi$, leaving $\bar{n}_\mathrm{a}^\mathrm{out}$ and $\bar{n}_\mathrm{r}^\mathrm{in}$ frequency-independent within the detection bandwidth
\begin{align}
\bar{n}_\mathrm{a}^\mathrm{out} &= \bar{n}_\mathrm{VTS} t_\mathrm{leak} + \bar{n}_\mathrm{ext} + \bar{n}_\mathrm{add}, \label{eq_nout}\\
\bar{n}_\mathrm{r}^\mathrm{in} &= \bar{n}_\mathrm{a}^\mathrm{out} t_\mathrm{loss} + \bar{n}_\mathrm{loss} (1-t_\mathrm{loss})\\
&= \bar{n}_\mathrm{para} t_\mathrm{loss} + (\bar{n}_\mathrm{ext} + \bar{n}_\mathrm{add}) t_\mathrm{loss}, \label{eq_nin} \\
\bar{n}_\mathrm{r}^\mathrm{eff} &= \frac{\kappa_\mathrm{a} \kappa_\mathrm{r,c}}{\kappa_\mathrm{r}^{2}} \bar{n}_\mathrm{r}^\mathrm{in}, \label{eq_neff}
\end{align}
in which $\bar{n}_\mathrm{para} \coloneqq \bar{n}_\mathrm{VTS} t_\mathrm{leak} + \bar{n}_\mathrm{loss} (1-t_\mathrm{loss})/ t_\mathrm{loss}$ denotes the parasitic white noise referred to the antenna input.
As shown in Fig.~\ref{fig_nid}(a), we operated JPC close to its full-conversion point, and measured $\bar{n}_\mathrm{r}^\mathrm{eff}$ as we ramped up the power of added white noise at room temperature.
Using Eqs.~(\ref{eq_nin}) and (\ref{eq_neff}), we thus obtain the linear coefficient between $P_\mathrm{add}$ and $\bar{n}_\mathrm{r}^\mathrm{eff}$, and convert $\bar{n}_\mathrm{r}^\mathrm{eff}$ to $\bar{n}_\mathrm{add}$ using $t_\mathrm{loss}=0.57$ (see Sec.~\ref{sec_cal}).
All the values of $\bar{n}_\mathrm{add}$ reported in the Main Text are calibrated using this method.

In Fig.~\ref{fig_nid}(b), we fixed $P_\mathrm{add}=2.5\ \si{mW}$ and swept the JPC-pump power.
The measured readout-mode populations fall on a rescaled JPC-conversion curve [blue in Fig.~\ref{fig_jpc}(a)], which shows that the JPC frequency conversion is noiseless in the sub-unit-photon regime.

\subsection{Dynamic range}

Here we give an estimate of the dynamic range of our radiometer. The detection limit of $\bar{n}_\mathrm{r}^\mathrm{eff}$ is set by $\delta \Gamma_\mathrm{2R} / \kappa_\mathrm{r} \sim 10^{-3}$, in which $\delta \Gamma_\mathrm{2R}$ is the fluctuation of the inherent qubit decoherence rate measured over 7 hours.
The upper bound of detectable $\bar{n}_\mathrm{r}^\mathrm{eff}$ is approximately $2\pi(f_\mathrm{q}^{ge}\!-\! f_\mathrm{q}^{ef})/\chi \sim 10^2$.
This restriction arises from the phenomenological observation that the maximum AC Stark shift of a transmon qubit ($\bar{n}_\mathrm{r} \chi$) appears to be limited by the qubit anharmonicity.
From these approximations, we estimate a dynamic range of the radiometer around $10^2 / 10^{-3} \sim 50$ dB.

\section{Theory of Colored-Noise-Induced Qubit Dephasing} \label{sec_color}

Previously, theories of qubit dephasing induced by coherent drives, white noise, and two-mode squeezed lights have been studied in Refs.~\cite{Gambetta2006}, \cite{Clerk2007}, and \cite{Silveri2016}, respectively.
However, in our experiment, the antenna noise incident on the radiometer is pulsed and frequency-dependent.
Here we present two approaches for predicting the qubit dephasing induced by pulsed Lorentzian thermal noise.
They quantitatively agree for the conditions of our experiment.
For simplicity, only the results of the first approach, which is analytical, are compared with the experimental data in the Main Text.

\subsection{Small-thermal-population approximation} \label{sec_approx}

In this subsection, we will derive an analytical expression for the average dephasing rate $\bar{\Gamma}_\mathrm{a}$ and the effective thermal population $\bar{n}_\mathrm{r}^\mathrm{eff}$ as a function of the various bath populations if the antenna radiation has a Lorentzian spectrum---the condition of our radiometry experiment.

First, consider an ideal situation where $\bar{n}_\mathrm{para}=0$ ($t_\mathrm{leak}=0$, $t_\mathrm{loss}=1$).
We use $\hat{a}$ and $\hat{b}$ to denote the antenna and the qubit-readout mode, respectively.
The Hamiltonians of the two subsystems are written as
\begin{align}
\frac{\hat{H}_a}{\hbar} &= \omega_\mathrm{a} \hat{a}^\dagger \hat{a} + \frac{\hat{H}_a^\mathrm{\kappa}}{\hbar} + \frac{\hat{H}_a^\mathrm{\mathrm{d}}}{\hbar}, \\
\frac{\hat{H}_b}{\hbar} &= \omega_\mathrm{r} \hat{b}^\dagger \hat{b} + \frac{\omega_\mathrm{q}^{ge} }{2} \hat{\sigma}_z - \frac{\chi }{2} \hat{b}^\dagger \hat{b} \hat{\sigma}_z + \frac{\hat{H}_b^\mathrm{\kappa}}{\hbar} + \frac{\hat{H}_b^\mathrm{\mathrm{d}}}{\hbar} .
\end{align}
In the above equations, $\omega_j = 2\pi f_j$ is the angular frequency of mode $j$; $\hat{\sigma}_z$ is the qubit-state operator, whose eigenvalue $\pm 1$ refers to the qubit being in its $e$ or $g$ state; Hamiltonians with superscripts ``$\kappa$'' and ``d'' formally describe dissipations and drives.

The Heisenberg--Langevin equations of the two cavity modes are
\begin{align}
\dot{\hat{a}}(t) &= - \!\left(\! i \omega_\mathrm{a} \!+\! \frac{\kappa_\mathrm{a}}{2} \! \right) \! \hat{a}(t) + \sqrt{\kappa_\mathrm{a,i}} \hat{A}_\mathrm{in,i}(t) + \sqrt{\kappa_\mathrm{a,c}} \hat{A}_\mathrm{in,c}(t), \label{eq_HLa} \\
\dot{\hat{b}}(t) &= - \!\left[ i \!\left( \!\omega_\mathrm{r} - \frac{\chi}{2} \hat{\sigma}_z \!\right)\! + \frac{\kappa_\mathrm{r}}{2}  \right] \hat{b}(t) \nonumber \\
&\qquad + \sqrt{\kappa_\mathrm{r,i}} \hat{B}_\mathrm{in,i}(t) + \sqrt{\kappa_\mathrm{r,c}} \hat{B}_\mathrm{in,c}(t). \label{eq_HLb}
\end{align}
Define the readout mode operator conditioned on the qubit state as
\begin{align}
\hat{b}_\sigma(t) \coloneqq \bra{\sigma} \hat{b} (t) \ket{\sigma},
\end{align}
in which $\sigma=\pm 1$ ($e$ or $g$). Here we disregard the energy relaxation of the qubit during the process, so that $\sigma$ is a constant of motion. We rewrite Eq.~(\ref{eq_HLb}) as
\begin{align}
\dot{\hat{b}}_\sigma(t) &= -\! \left(\! i \omega_\mathrm{r}^\sigma \!+\! \frac{\kappa_\mathrm{r}}{2} \! \right)\! \hat{b}_\sigma(t) + \sqrt{\kappa_\mathrm{r,i}} \hat{B}_\mathrm{in,i}(t) + \sqrt{\kappa_\mathrm{r,c}} \hat{B}_\mathrm{in,c}(t), \label{eq_HLb_sigma}
\end{align}
where $\omega_\mathrm{r}^\sigma = \omega_\mathrm{r} - \sigma \chi /2$ is the qubit-state-dependent readout-cavity frequency.

The input--output relations \cite{Clerk2010} conditioned on the qubit state are
\begin{align} \label{eq_io}
\hat{A}_\mathrm{out,c} &= \hat{A}_\mathrm{in,c} - \sqrt{\kappa_\mathrm{a,c}}\, \hat{a} = \hat{B}_\mathrm{in,c} ,\\
\hat{B}_\mathrm{out,c}^\sigma&= \hat{B}_\mathrm{in,c} - \sqrt{\kappa_\mathrm{r,c}}\, \hat{b}_\sigma.
\end{align}
The input-mode operators $\hat{A}_\mathrm{in,i}$ and $\hat{A}_\mathrm{in,c}$ ($\hat{B}_\mathrm{in,i}$ and $\hat{B}_\mathrm{in,c}$) denote the thermal fluctuations of the internal and external baths of the antenna resonator (qubit-readout cavity).
$\hat{B}_\mathrm{in,i}$ can be disregarded as we are calculating only the qubit dephasing induced by the antenna radiation incident on the qubit-readout cavity.
We write the time-correlation functions of the stationary white thermal baths
\begin{align}
\braket{\hat{A}_\mathrm{in,i}^\dagger(t) \hat{A}_\mathrm{in,i}(t')} &= \bar{n}_\mathrm{VTS} \delta(t-t'),\\
\braket{\hat{A}_\mathrm{in,i}(t) \hat{A}_\mathrm{in,i}^\dagger(t')} &= (\bar{n}_\mathrm{VTS}+1) \delta(t-t'),\\
\braket{\hat{A}_\mathrm{in,c}^\dagger(t) \hat{A}_\mathrm{in,c}(t')} &= (\bar{n}_\mathrm{ext} + \bar{n}_\mathrm{add}) \delta(t-t'),\\
\braket{\hat{A}_\mathrm{in,c}(t) \hat{A}_\mathrm{in,c}^\dagger(t')} &= (\bar{n}_\mathrm{ext} + \bar{n}_\mathrm{add} + 1) \delta(t-t'),
\end{align}
and their spectral densities
\begin{align}
\braket{\hat{A}_\mathrm{in,i}^\dagger \hat{A}_\mathrm{in,i}}[\omega] &= \bar{n}_\mathrm{VTS},\\
\braket{\hat{A}_\mathrm{in,i} \hat{A}_\mathrm{in,i}^\dagger}[\omega] &= \bar{n}_\mathrm{VTS}+1 ,\\
\braket{\hat{A}_\mathrm{in,c}^\dagger \hat{A}_\mathrm{in,c}}[\omega] &= \bar{n}_\mathrm{ext} + \bar{n}_\mathrm{add} ,\\
\braket{\hat{A}_\mathrm{in,c} \hat{A}_\mathrm{in,c}^\dagger}[\omega] &= \bar{n}_\mathrm{ext} + \bar{n}_\mathrm{add} + 1,
\end{align}
which are related via
\begin{align}
\braket{\hat{F}_1 \hat{F}_2}[\omega] = \int_{-\infty}^{\infty} \braket{\hat{F}_1(t+\tau) \hat{F}_2(t)} e^{i\omega\tau}\, \mathrm{d}\tau.
\end{align}
Fourier-transform Eq.~(\ref{eq_HLa}) into the frequency domain:
\begin{align}
\left[ \frac{\kappa_\mathrm{a}}{2} - i (\omega-\omega_\mathrm{a}) \right] \hat{a}[f] = \sqrt{\kappa_\mathrm{a,i}} \hat{A}_\mathrm{in,i}[f] + \sqrt{\kappa_\mathrm{a,c}} \hat{A}_\mathrm{in,c}[f]. \label{eq_FT}
\end{align}
From Eqs.~(\ref{eq_io}) and (\ref{eq_FT}), we obtain
\begin{align} \label{eq_lrz}
& \quad \braket{\hat{B}_\mathrm{in, c}^\dagger \hat{B}_\mathrm{in,c}}[\omega] = \braket{\hat{A}_\mathrm{out, c}^\dagger \hat{A}_\mathrm{out,i}}[\omega] \nonumber\\
&= \bar{n}_\mathrm{VTS} t_\mathrm{a}[\omega \!-\!\omega_\mathrm{a}] + (\bar{n}_\mathrm{ext} \!+\! \bar{n}_\mathrm{add})(1 \!-\! t_\mathrm{a}[\omega\!-\!\omega_\mathrm{a}] )\\
&= (\bar{n}_\mathrm{VTS}\!-\!\bar{n}_\mathrm{ext} \!-\! \bar{n}_\mathrm{add}) t_\mathrm{a}[\omega \!-\!\omega_\mathrm{a}] + \bar{n}_\mathrm{ext} + \bar{n}_\mathrm{add} ,
\end{align}
in which $t_\mathrm{a}$ is the Lorentzian transmission function
\begin{equation}
t_\mathrm{a}[\omega\!-\!\omega_\mathrm{a}]
= \frac{4 \kappa_\mathrm{a,i} \kappa_\mathrm{a,c}}{ \kappa_\mathrm{a}^2 + 4(\omega\!-\!\omega_\mathrm{a})^2}
= \frac{4 \gamma (1\!-\!\gamma)}{1 + 4(\omega\!-\!\omega_\mathrm{a})^2 \!/ \!\kappa_\mathrm{a}^2 },
\end{equation}
where $\gamma = \kappa_\mathrm{a,i}/\kappa_\mathrm{a}$.

We now set $\bar{n}_\mathrm{ext} = \bar{n}_\mathrm{add} = 0$ in Eq.~(\ref{eq_lrz}) and focus on the effect of transmitted noise from the internal bath---the term proportional to $t_\mathrm{a}$. In the time domain, the correlation functions of $\hat{B}_\mathrm{in, c}^\dagger$ and $\hat{B}_\mathrm{in,c}$ have bilateral exponential forms. We define
\begin{align} \label{eq_lap}
L(t,t') \coloneqq \kappa_\mathrm{a} \gamma (1\!-\!\gamma) e^{-\frac{\kappa_\mathrm{a}}{2}|t-t'|} e^{i\omega_\mathrm{a}(t-t')},
\end{align}
and can thus write
\begin{align}
\braket{\hat{B}_\mathrm{in, c}^\dagger(t) \hat{B}_\mathrm{in,c}(t')} &=
\bar{n}_\mathrm{VTS} L(t,t'), \\
\braket{\hat{B}_\mathrm{in, c}(t) \hat{B}_\mathrm{in,c}^\dagger(t')} &=
(\bar{n}_\mathrm{VTS}+1) L^*(t,t'), \\
\braket{\hat{B}_\mathrm{in, c}(t) \hat{B}_\mathrm{in,c}(t')} &=\braket{\hat{B}_\mathrm{in, c}^\dagger(t) \hat{B}_\mathrm{in,c}^\dagger(t')} = 0.
\end{align}

One can consider the input colored thermal noise as an incoherent mixture of coherent states. For a coherent-state input, the total cavity-induced qubit dephasing, namely, the off-diagonal element of the qubit density matrix after the cavity reaches a steady state is given by $e^{-\int \Gamma_\mathrm{coh}(t) \, \mathrm{d}t}$, where \cite{Gambetta2006,Silveri2016}
\begin{align} \label{eq_coh}
\Gamma_\mathrm{coh}(t) = \frac{\kappa_\mathrm{r}}{2} \left| \beta_g(t) - \beta_e(t) \right|^2,
\end{align}
in which $\beta_g$ and $\beta_e$ are the (complex) coherent-state amplitudes of the readout cavity given the qubit being in $\ket{g}$ and $\ket{e}$.
In the case of thermal noise input, the dephasing can be obtained by averaging over all possible realizations of the input coherent state $\braket{e^{-\int \Gamma_\mathrm{coh}(t) \, \mathrm{d}t}}$. One can simplify this expression through the cumulant expansion of averages, that is,
\begin{align} \label{eq_cumu}
\left\langle e^{-\!\int \Gamma_\mathrm{coh}(t) \, \mathrm{d}t} \right\rangle = e^{- \! \int \braket{\Gamma_\mathrm{coh}(t)} \mathrm{d}t \,+\, \frac{1}{2} \! \iint \braket{\Gamma_\mathrm{coh}(t)\Gamma_\mathrm{coh}(t')}\, \mathrm{d}t\mathrm{d}t'+\cdots }
\end{align}
Higher-order cumulants can be neglected in the limit of small input thermal photon number, namely, $\gamma \bar{n}_\mathrm{VTS} \ll 1$.
In this approximation, the total dephasing rate of the qubit
is given by
\begin{align}
\Gamma_\mathrm{a}(t) = \braket{\Gamma_\mathrm{coh}(t)} = \frac{\kappa_\mathrm{r}}{2} \left\langle \left|\beta_g(t) - \beta_e(t)\right|^2 \right\rangle.
\end{align}
In the Heisenberg picture, the above equation can be written as
\begin{align}
\Gamma_\mathrm{a}(t) &=
\frac{\kappa_\mathrm{r}}{2}\! \braket{\hat{D}^\dagger(t) \hat{D}(t)} , \label{eq_DdD}
\end{align}
in which
\begin{align}
\hat{D}(t) & \coloneqq \hat{b}_g(t) - \hat{b}_e(t).
\end{align}
Consequently, the average dephasing rate defined in the Main Text is

\begin{align}
\bar{\Gamma}_\mathrm{a} &= \frac{\kappa_\mathrm{r}}{2 \tau_\mathrm{p}} \int_0^\tau \braket{\hat{D}^\dagger(t) \hat{D}(t)} \mathrm{d}t \label{eq_intD} \\
& = \frac{\kappa_\mathrm{r}}{2 \tau_\mathrm{p}}  \sum_{\sigma_1, \sigma_2} \mathrm{sgn}[\sigma_1\sigma_2] \int_0^\tau \braket{\hat{b}_{\sigma_1}^\dagger(t) \hat{b}_{\sigma_2}(t)}  \mathrm{d}t,
\end{align}
in which $\tau$ is the separation between the two $\pi/2$ qubit pulses and $\tau_\mathrm{p}$ is the duration of the pulsed JPC pump, as indicated in Fig.~2(a). To calculate the right-hand side of Eq.~(\ref{eq_intD}), we solve the dynamics of the readout mode operators from Eq.~(\ref{eq_HLb_sigma}):
\begin{equation}
\begin{split}
\hat{b}_\sigma(t) &= \hat{b}_\sigma(0) e^{-\left(i\omega_\mathrm{r}^\sigma + \frac{\kappa_\mathrm{r}}{2} \right)t}  \\
&\quad +\sqrt{\kappa_\mathrm{r,c}} \int_0^\tau \hat{B}_\mathrm{in,c} e^{-\left(i\omega_\mathrm{r}^\sigma + \frac{\kappa_\mathrm{r}}{2} \right)(t-t')}\, \mathrm{d}t'.
\end{split}
\end{equation}
In our experiment, before the JPC pump is applied, the readout cavity is free from the antenna radiation,
\begin{align}
\braket{\hat{b}_{\sigma_1}^\dagger(0) \hat{b}_{\sigma_2}(0)} = 0.
\end{align}
Therefore,
\begin{equation}
\begin{split}
&\braket{\hat{b}_{\sigma_1}^\dagger(t) \hat{b}_{\sigma_2}(t)} = \kappa_\mathrm{r,c} e^{-\kappa_\mathrm{r}t} e^{i \left( \omega_\mathrm{r}^{\sigma_1} - \omega_\mathrm{r}^{\sigma_2} \right)t} \\
&\times \!\! \iint_0^\tau \!\! \braket{\hat{B}_\mathrm{in, c}^\dagger(t') \hat{B}_\mathrm{in,c}(t'')} e^{\frac{\kappa_\mathrm{r}}{2}(t'+t'')} e^{i \left( \omega_\mathrm{r}^{\sigma_2} t'' - \omega_\mathrm{r}^{\sigma_1} t' \right)} \, \mathrm{d}t' \mathrm{d}t'' \!,
\end{split}
\end{equation}
in which
\begin{equation}
\begin{split}
\braket{\hat{B}_\mathrm{in, c}^\dagger(t') \hat{B}_\mathrm{in,c}(t'')}=
   \begin{cases}
      \bar{n}_\mathrm{VTS} L (t',t''), & 0<t',t''<\tau_\mathrm{p}, \\
      0, & \mathrm{otherwise}.
   \end{cases}
\end{split}
\end{equation}
This condition is due to the fact that the JPC is pulsed. Evaluating these integrals for $t<\tau_\mathrm{p}$ yields
\begin{align}
\braket{\hat{b}_{g}^\dagger(t) \hat{b}_{g}(t) } &= N(t, \kappa_\mathrm{r}, \Delta_\mathrm{a} \!-\! \chi/2),\\
\braket{\hat{b}_{e}^\dagger(t) \hat{b}_{e}(t) } &= N(t, \kappa_\mathrm{r}, \Delta_\mathrm{a} \!+\! \chi/2),\\
\braket{\hat{b}_{g}^\dagger(t) \hat{b}_{e}(t) } &= N(t, \kappa_\mathrm{r} \!-\! i\chi, \Delta_\mathrm{a} ),\\
\braket{\hat{b}_{e}^\dagger(t) \hat{b}_{g}(t) } &= N(t, \kappa_\mathrm{r} \!+\! i\chi, \Delta_\mathrm{a} ),
\end{align}
in which $\Delta_\mathrm{a} = \omega_\mathrm{a} - \omega_\mathrm{r} - \omega_\mathrm{p}$, and
\begin{equation}
\begin{split}
&N(t,\kappa,\Delta) = \bar{n}_\mathrm{VTS} \kappa_\mathrm{a} \kappa_\mathrm{r,c} \gamma (1-\gamma) \\
& \quad \times \! \left\{ \!\frac{\left(\frac{\kappa_\mathrm{a}}{\kappa}\!-\! 1 \right) \!\! \left(1 \!-\! e^{-\kappa t} \right)}{\Delta^2 + (\frac{\kappa_\mathrm{a}-\kappa}{2})^2} + \!\left[\frac{1 - e^{\left( i\Delta - \frac{ \kappa_\mathrm{a}+\kappa}{2} \right) t}}{\left(\Delta \! + \! i\frac{\kappa_\mathrm{a}}{2}\right)^2 \!+\! \frac{\kappa^2}{4}} + \mathrm{c.c.} \right] \! \right\} \!.
\end{split}
\end{equation}
When $t>\tau_\mathrm{p}$,
\begin{align}
\braket{\hat{b}_{g}^\dagger(t) \hat{b}_{g}(t) } &= \braket{\hat{b}_{g}^\dagger(\tau_\mathrm{p}) \hat{b}_{g}(\tau_\mathrm{p}) } e^{-\kappa_\mathrm{r}(t-\tau_\mathrm{p})}, \label{eq_bgbg} \\
\braket{\hat{b}_{e}^\dagger(t) \hat{b}_{e}(t) } &= \braket{\hat{b}_{e}^\dagger(\tau_\mathrm{p}) \hat{b}_{e}(\tau_\mathrm{p}) } e^{-\kappa_\mathrm{r}(t-\tau_\mathrm{p})}, \\
\braket{\hat{b}_{g}^\dagger(t) \hat{b}_{e}(t) } &= \braket{\hat{b}_{g}^\dagger(\tau_\mathrm{e}) \hat{b}_{e}(\tau_\mathrm{p}) } e^{-(\kappa_\mathrm{r}-i\chi) (t-\tau_\mathrm{p})}, \\
\braket{\hat{b}_{e}^\dagger(t) \hat{b}_{g}(t) } &= \braket{\hat{b}_{e}^\dagger(\tau_\mathrm{p}) \hat{b}_{g}(\tau_\mathrm{p}) } e^{-(\kappa_\mathrm{r}+i\chi) (t-\tau_\mathrm{p})}.  \label{eq_bebg}
\end{align}
We plug these expressions into Eq.~(\ref{eq_intD}) and evaluate $\bar{\Gamma}_\mathrm{a}$, which is linked to $\bar{n}_\mathrm{r}^\mathrm{eff}$ via Eq.~(\ref{eq_white}). Particularly, when $\bar{n}_\mathrm{r}^\mathrm{eff}\ll 1$,
\begin{equation}
\begin{split}
\bar{n}_\mathrm{r}^\mathrm{eff} = \frac{\chi^2\! +\! \kappa_\mathrm{r}^2}{\chi^2\kappa_\mathrm{r}} \bar{\Gamma}_\mathrm{a}. \label{eq_dephs}
\end{split}
\end{equation}

\begin{figure*}[t]
\includegraphics[width=\textwidth]{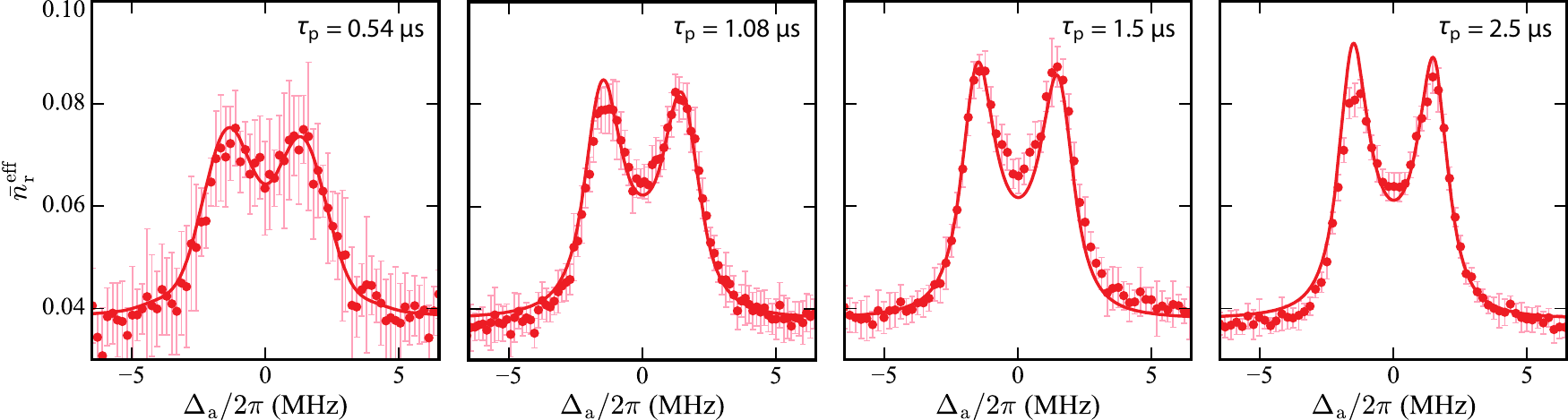}
\caption{Qubit-dephasing spectra with different JPC-pump durations and $\bar{n}_\mathrm{add}=0$.
Parameters of the theoretical curves are taken from Table~I:
$\bar{n}_\mathrm{VTS}=1.59$ ($T_\mathrm{VTS}=1.03\ \si{K}$),
$\bar{n}_\mathrm{ext}=0.014$,
$\bar{n}_\mathrm{loss}=0.09$,
$t_\mathrm{leak}=0.046$,
and $t_\mathrm{loss}=0.57$.
}
\label{fig_tp}
\end{figure*}

The dimensionless function $\eta_\mathrm{a}$ appearing in Eq.~(4) and plotted in Fig.~4(b) is therefore defined as
\begin{equation}
\begin{split}
\eta_\mathrm{a}[\Delta_\mathrm{a},\tau_\mathrm{p}] \coloneqq \frac{\bar{n}_\mathrm{r}^\mathrm{eff} [\Delta_\mathrm{a},\tau_\mathrm{p}] }{\bar{n}_\mathrm{VTS}} \frac{\kappa_\mathrm{r}^{2}}{\kappa_\mathrm{r,c} \kappa_\mathrm{a}},
\end{split}
\end{equation}
which approaches zero when $|\Delta_\mathrm{a}|\gg \chi,\,\kappa_\mathrm{r}$.
Similarly, if we set $\bar{n}_\mathrm{VTS} = 0$ in Eq.~(\ref{eq_lrz}) and only study the qubit dephasing induced by the reflected thermal noise from the external bath, then we can get
\begin{equation}
\begin{split}
\bar{n}_\mathrm{r}^\mathrm{eff} [\Delta_\mathrm{a},\tau_\mathrm{p}]  = \frac{\kappa_\mathrm{a} \kappa_\mathrm{r,c}}{\kappa_\mathrm{r}^{2}} \!\left(\bar{n}_\mathrm{ext}\! +\! \bar{n}_\mathrm{add} \right)\! \left(1 \!-\! \eta_\mathrm{a}[\Delta_\mathrm{a},\tau_\mathrm{p}] \right) \! .
\end{split}
\end{equation}
Taking the transmitted, reflected, and white parasitic noise all into consideration, we thus obtain Eq.~(4)---the response function of our quantum radiometer, which is the theoretical basis for the noise analysis.

The results of this analytical approach are plotted in Fig.~4(b) of the Main Text to be compared with the experimental data.
Furthermore, we show in Fig.~\ref{fig_tp} for qubit-dephasing spectra ($\bar{n}_\mathrm{r}^\mathrm{eff}$--$f_\mathrm{a}$ relations) measured with different choices of JPC-pump duration $\tau_\mathrm{p}$, and plotted the theoretical predictions according to Eq.~(4).
In the theoretical model, the bath temperatures and $t_\mathrm{leak}$ are set to be their average values according to the radiometry results shown in Table I, and we choose $t_\mathrm{loss}=0.57$ (within its 1-$\sigma$ range: $0.52\pm0.06$) to yield the best experiment--theory agreement.
As expected from time--frequency duality, the value of $\tau_\mathrm{p}$ limits the frequency resolution to $\sim 1/\tau_\mathrm{p}$.
With $\tau_\mathrm{p}=2.5\ \si{\us}$, the peak on the qubit-dephasing spectrum at $\Delta_\mathrm{a} = -\chi/2$ is visibly lower than the one at $\Delta_\mathrm{a} = \chi/2$, which is due to the qubit-relaxation during the Ramsey experiment.

The main approximation made in deriving the expressions for the total qubit-dephasing rate is truncating the cumulant expansion after the first-order term.
This approximation breaks down as the number of input thermal photons $\gamma \bar{n}_\mathrm{VTS}$ increases.
For large numbers of input thermal photons, Eq.~(\ref{eq_cumu}) overestimates the dephasing rate, because the second-order correction in the cumulant expansion is subtracted from the leading-order term.
In order to verify our approximation, we compare the results in this subsection with exact numerical solutions to the master equation.

\subsection{Beyond small-thermal-population approximation}

In this subsection, we solve the master equation for cascaded systems using the $P$-function method.

To begin with, we write the density operator of the radiometer in the form
\begin{align} \label{eq_des_rho}
\hat{\rho}(t) = \sum_{\sigma_1, \sigma_2 = g, e} \ket{\sigma_1}\!\bra{\sigma_2}\otimes \hat{\rho}_{\sigma_1 \sigma_2}(t),
\end{align}
in which $\ket{\sigma_{1,2}}$ refers to the qubit state, and $\hat{\rho}_{\sigma_1 \sigma_2}$ is the joint density operator of the \emph{antenna} plus the \emph{readout cavity} conditioned on the qubit state.

First, we set $\bar{n}_\mathrm{ext} = \bar{n}_\mathrm{add} = 0$ and study the qubit dephasing induced by the transmitted noise from the internal bath with a thermal population $\bar{n}_\mathrm{VTS}$.
The cascaded master equation of the system, written in the frame rotating at the antenna frequency $\omega_\mathrm{a}$, is
\begin{align}
\dot{\hat{\rho}} &= \frac{1}{i\hbar}\! \left[\hat{\tilde{H}}, \hat{\rho} \right] \! + \mathcal{D}\! \left[\sqrt{\kappa_\mathrm{r,c}}\htb \right]\! \hat{\rho} + \mathcal{D} \! \left[ \sqrt{\kappa_\mathrm{r,i}}\htb \right] \hat{\rho} + \mathcal{D}\! \left[\sqrt{\kappa_\mathrm{a,c}}\hta \right] \! \hat{\rho} \nonumber \\
& \quad + \mathcal{D} \! \left[\sqrt{(\bar{n}_\mathrm{VTS} \! + \! 1) \kappa_\mathrm{a,i} } \hat{a} \right]\! \hat{\rho} + \mathcal{D} \! \left[\sqrt{ \bar{n}_\mathrm{VTS} \kappa_\mathrm{a,i} } \hta^\dag \right] \! \hat{\rho} \nonumber \\
& \quad - \sqrt{ \kappa_\mathrm{a,c} \kappa_\mathrm{r,c} } \left( \left[ \htb^\dag,\hta\hat{\rho} \right]+ \left[\hat{\rho}\hta^\dag,\htb \right] \right),
\label{eq_me1}
\end{align}
in which
\begin{align}
\frac{\hat{\tilde{H}}}{\hbar} = - \left(\Delta_\mathrm{a} + \frac{\chi  }{2}\hat{\sigma}_z \right) \hat{b}^\dagger \hat{b} + \frac{\omega_{ge}}{2} \hat{\sigma}_z
\end{align}
is the dispersive Hamiltonian in the rotating frame, and
\begin{align}
\mathcal{D} \! \left[\hat{A} \right]\! \hat{\rho} = \hat{A} \hat{\rho} \hat{A}^\dagger - \frac{1}{2} \left\{\hat{A}^\dagger \hat{A}, \hat{\rho} \right\}
\end{align}
is the Lindblad superoperator.
Moreover, we use a time-dependent coupling rate
\begin{equation}
\kappa_\mathrm{r,c}(t) = \kappa_\mathrm{r,c} \left[\Theta(0) - \Theta(\tau_\mathrm{p})\right]
\end{equation}
to model the pulsed operation of the JPC switch,
where $\Theta(t)$ is the Heaviside step function.
In our Ramsey experiment, the qubit is initialized in $(\ket{g} + \ket{e})/\sqrt{2}$ and the readout cavity is initialized in vacuum.
At $\tau=0$, we have $\mathrm{Tr} [ \hat{\rho}_{ge}(0) ] =1/2$. The information about qubit dephasing is therefore carried by $\hat{\rho}_{ge}$ (or $\hat{\rho}_{eg}$):
\begin{align}\label{eq_tr}
\frac{A_\mathrm{on}}{A_\mathrm{off}} = 2 \left|\mathrm{Tr} \left[ \hat{\rho}_{ge}(\tau) \right] \right|.
\end{align}

\begin{figure}[b]
\includegraphics[width=\columnwidth]{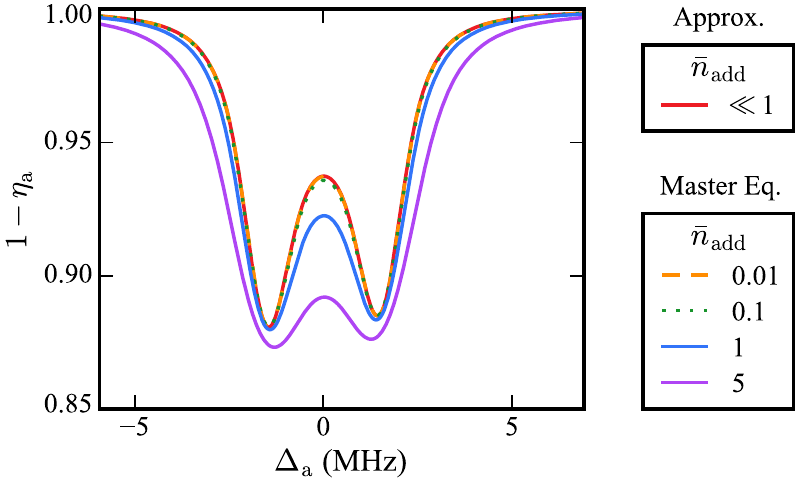}
\caption{Predictions of the dimensionless detector response function $\eta_\mathrm{a}$ by the approximate analytical theory compared with the numerical solutions of the master equations given different added photon numbers.
Here we set $\bar{n}_\mathrm{ext}=0$.
Other experimental parameters are chosen to be the same as those used in Fig.~4 of the Main Text.
}
\label{fig_eta_ME}
\end{figure}

From Eqs.~(\ref{eq_des_rho}) and (\ref{eq_me1}), we obtain the equation of motion for $\hat{\rho}_{ge}$:
\begin{align}
\dot{\hat{\rho}}_{ge} &= i \Delta_\mathrm{a}\! \left[\htb^\dagger \htb, \hat{\rho}_{ge} \right]  - \frac{i \chi}{2} \left\{\htb^\dagger \htb, \hat{\rho}_{ge} \right\} \nonumber \\
& \quad + \mathcal{D}\! \left[\sqrt{\kappa_\mathrm{r,c}}\htb \right]\! \hat{\rho}_{ge} + \mathcal{D} \! \left[ \sqrt{\kappa_\mathrm{r,i}}\htb \right] \hat{\rho}_{ge} + \mathcal{D}\! \left[\sqrt{\kappa_\mathrm{a,c}}\hta \right] \! \hat{\rho}_{ge} \nonumber \\
& \quad + \mathcal{D} \! \left[\sqrt{(\bar{n}_\mathrm{VTS} \! + \! 1) \kappa_\mathrm{a,i} } \hat{a} \right]\! \hat{\rho}_{ge} + \mathcal{D} \! \left[\sqrt{ \bar{n}_\mathrm{VTS} \kappa_\mathrm{a,i} } \hta^\dag \right] \! \hat{\rho}_{ge} \nonumber \\
& \quad - \sqrt{ \kappa_\mathrm{a,c} \kappa_\mathrm{r,c} } \left( \left[ \htb^\dag, \hta \hat{\rho}_{ge} \right]+ \left[\hat{\rho}_{ge} \hta^\dag, \htb \right] \right),
\label{eq_me2}
\end{align}
To solve Eq.~(\ref{eq_me2}), we use the $P$-representation \cite{Walls2008} of the two bosonic modes ($\alpha,\alpha^*\leftrightarrow \hta,\hta^\dag$ and $\beta,\beta^*\leftrightarrow\htb,\htb^\dag$):
\begin{align}
\hat{\rho}_{ge}(t)&=\int P(\alpha,\alpha^*\!,\beta,\beta^*\!,t) \ket{\alpha,\beta}\!\bra{\alpha,\beta}\, \mathrm{d}^2\alpha \, \mathrm{d}^2\beta,
\end{align}
and write the Fokker--Planck equation \cite{Scully1997} for $P$:
\begin{align}
\frac{\partial P}{\partial t} &= \left[ \frac{\kappa_\mathrm{r} + i \chi}{2} - i\Delta_\mathrm{a} \right] \frac{\partial (\beta P)}{\partial \beta} \nonumber \\
& \quad + \left[\frac{\kappa_\mathrm{r} + i \chi}{2} + i\Delta_\mathrm{a} \right] \frac{\partial (\beta^* P)}{\partial \beta^*}-i\chi|\beta|^2P \nonumber\\
& \quad + \frac{\kappa_\mathrm{a}}{2} \left[\frac{\partial (\alpha P)}{\partial \alpha}+\frac{\partial (\alpha^* P)}{\partial \alpha^*}\right] + \bar{n}_\mathrm{VTS} \kappa_\mathrm{a,i} \frac{\partial^2 P}{\partial\alpha\partial\alpha^*} \nonumber \\
& \quad + \sqrt{\kappa_\mathrm{a,c} \kappa_\mathrm{r,c}} \left(\alpha\frac{\partial P}{\partial\beta}+\alpha^*\frac{\partial P}{\partial\beta^*}\right),
\label{FP}
\end{align}
whose initial condition is
\begin{equation} \label{eq_init}
P(\alpha,\alpha^*\!,\beta,\beta^*\!,0) = \frac{1}{\pi \gamma \bar{n}_\mathrm{VTS}} e^{-\frac{|\alpha|^2}{\gamma\bar{n}_\mathrm{VTS} }} \delta(\beta) \delta(\beta^*).
\end{equation}
Namely, the antenna resonator is in a thermal state with average photon population $\gamma \bar{n}_\mathrm{VTS}$, and the qubit readout cavity is in vacuum.

We write the solution for $P$ in the form of a Gaussian ansatz:
\begin{align}
P(\alpha,\alpha^*\!,\beta,\beta^*\!,t)\!=\!E(t)e^{-A(t)|\alpha|^2-B(t)|\beta|^2-C(t)\alpha\beta^*-D(t)\beta\alpha^*}\!,
\end{align}
and obtain the following ordinary differential equations for its coefficients:
\begin{align}
\dot{A} &= \kappa_\mathrm{a} A - \bar{n}_\mathrm{VTS} \kappa_\mathrm{a,i} A^2 + \sqrt{\kappa_\mathrm{a,c} \kappa_\mathrm{r,c}}(C+D) , \\
\dot{B} &= (\kappa_\mathrm{r} + i\chi)B - \bar{n}_\mathrm{VTS} \kappa_\mathrm{a,i}CD + i\chi , \\
\dot{C} &= \!\left(\! \frac{\kappa_\mathrm{a} \!+\! \kappa_\mathrm{r} \!+\! i\chi}{2} +  i\Delta_\mathrm{a} - \bar{n}_\mathrm{VTS} \kappa_\mathrm{a,i} A\! \right)\!C + \sqrt{\kappa_\mathrm{a,c} \kappa_\mathrm{r,c}} B, \\
\dot{D} &= \!\left(\! \frac{\kappa_\mathrm{a} \!+\! \kappa_\mathrm{r} \!+\! i\chi}{2} - i \Delta_\mathrm{a} - \bar{n}_\mathrm{VTS} \kappa_\mathrm{a,i} A\!\right)\!D  + \sqrt{\kappa_\mathrm{a,c} \kappa_\mathrm{r,c}} B, \\
\dot{E}& = \!\left(\kappa_\mathrm{a} + \kappa_\mathrm{r} + i\chi - \bar{n}_\mathrm{VTS} \kappa_\mathrm{a,i} A \right)\! E.
\label{difff}
\end{align}
Note that the set of differential equations above is closed and the Gaussian ansatz is therefore exact.
These equations can be numerically solved with the initial condition written in Eq.~(\ref{eq_init}).
Finally, the qubit dephasing can be calculated by integrating the $P$-function,
\begin{align}
\frac{A_\mathrm{on}}{A_\mathrm{off}} = 2 \left|\int P(\alpha,\alpha^*,\beta,\beta^*,\tau) \, \mathrm{d}^2\alpha \, \mathrm{d}^2\beta \right|,
\end{align}
The detector response function $\eta_\mathrm{a}$ can be subsequently obtained.

Similarly, we can set $\bar{n}_\mathrm{VTS} = 0$ and study the qubit dephasing induced by the reflected noise from the external bath with the thermal population $\bar{n}_\mathrm{ext}+\bar{n}_\mathrm{add}$.
According to the reasoning in the previous subsection, this calculation should yield the same $\eta_\mathrm{a}$ function.

In Fig.~\ref{fig_eta_ME}, we set $\bar{n}_\mathrm{ext}=0$ and compare the predictions of $\eta_\mathrm{a}$ by numerically solving the master equation at different values of $\bar{n}_\mathrm{add}$ with the outcomes of the approximate analytical theory introduced in Sec.~\ref{sec_approx}.
We find that these two approaches are equivalent when $\bar{n}_\mathrm{add} \ll 1$.
However, when $\bar{n}_\mathrm{add} \gtrsim 1$ where photon bunching of thermal radiation cannot be neglected, the predictions of the approximate theory, in which qubit dephasing rate is always proportional to $\bar{n}_\mathrm{add}$, deviate from the master-equation results.

Finally, we note that the cascaded master equation Eq.~(\ref{eq_me1}) is equivalent to a set of optical Bloch equations that describes the dynamics of $\langle \hat{\sigma}_+\rangle$, $\langle \hta^\dag \hta \hat{\sigma}_+ \rangle$, $\langle \htb^\dag \htb \hat{\sigma}_+ \rangle$, $\langle \htb^\dag \hta \hat{\sigma}_+ \rangle$, and $\langle \hta^\dag \htb \hat{\sigma}_+ \rangle$.
In the limit of small $\gamma \bar{n}_\mathrm{VTS}$, the same approximation allowing us to neglect the higher-order cumulants in Sec.~\ref{sec_approx} leaves these optical Bloch-equations to be a closed set of differential equations, which can be analytical solved using the Laplace transform in the limit of large $\chi$.
For the conditions of our experiment, the predictions of optical Bloch equations quantitatively agree with those of the two approaches introduced in this section.

\section{Radiometer Calibration} \label{sec_cal}

\begin{figure}[b]
\includegraphics[width=\columnwidth]{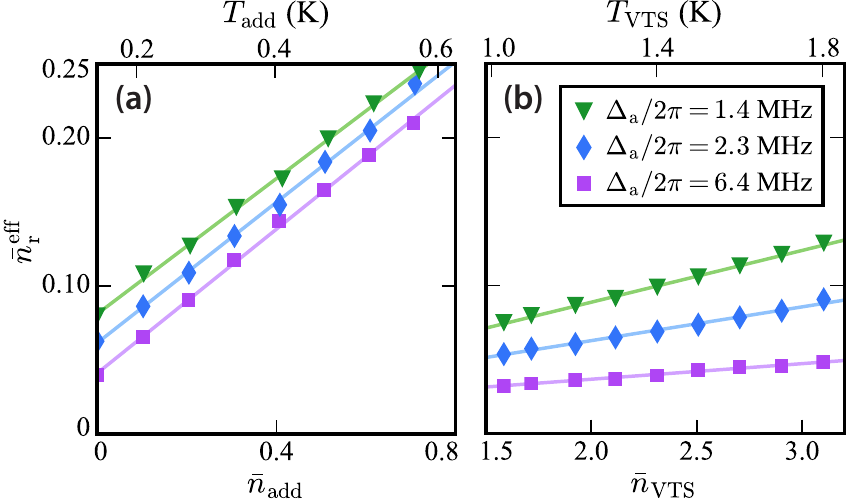}
\caption{Experimental data and linear fits of $\bar{n}_\mathrm{r}^\mathrm{eff}$ at different \textbf{(a)} VTS temperatures and \textbf{(b)} added white-noise powers.
In \textbf{(a)}, $T_\mathrm{add} = 0$ ($\bar{n}_\mathrm{add} = 0$).
In \textbf{(b)}, $T_\mathrm{VTS}=1.03\ \si{\K}$ ($\bar{n}_\mathrm{VTS} = 1.59$).
For all data in this figure, $\tau_\mathrm{p}=1.08\ \si{\us}$.}
\label{fig_vts}
\end{figure}

We calibrated our radiometer by measuring qubit-dephasing spectra while varying the thermal populations of the external and internal antenna baths.
To understand the procedure, we write Eq.~(4) again,
\begin{equation} \label{eq_color}
\begin{split}
& \bar{n}_\mathrm{r}^\mathrm{eff} [\Delta_\mathrm{a},\tau_\mathrm{p}]  = \kappa_\mathrm{r}^{-2} \kappa_\mathrm{a} \kappa_\mathrm{r,c} \!\left\{ \bar{n}_\mathrm{loss} (1\!-\!t_\mathrm{loss}) + \bar{n}_\mathrm{VTS} t_\mathrm{leak} t_\mathrm{loss} \right. \\
& + \left. \bar{n}_\mathrm{VTS} t_\mathrm{loss} \eta_\mathrm{a} [\Delta_\mathrm{a},\tau_\mathrm{p}] + \!\left(\bar{n}_\mathrm{ext} \!+\! \bar{n}_\mathrm{add} \right)\! t_\mathrm{loss}\! \left(1\!-\!\eta_\mathrm{a} [\Delta_\mathrm{a},\tau_\mathrm{p}]\right) \! \right\} \! ,
\end{split}
\end{equation}
in which $\Delta_\mathrm{a}$, $\tau_\mathrm{p}$, $\bar{n}_\mathrm{add}$, and $\bar{n}_\mathrm{VTS}$ are tunable parameters in our experiment, while $\bar{n}_\mathrm{ext}$, $\bar{n}_\mathrm{loss}$, $t_\mathrm{loss}$, and $t_\mathrm{leak}$ are to be extracted.
We chose $\tau_\mathrm{p}=1.08\ \si{\us}$ in the following measurements.
When $|\Delta_\mathrm{a}|\gg \chi$, Eq.~(\ref{eq_color}) is reduced to Eqs.~(\ref{eq_nout})--(\ref{eq_neff}) on white-noise-induced qubit dephasing,
\begin{equation} \label{eq_white2}
\begin{split}
\bar{n}_\mathrm{r}^\mathrm{eff} [\infty] & = \frac{\kappa_\mathrm{a} \kappa_\mathrm{r,c}}{\kappa_\mathrm{r}^{2}} \left\{ \bar{n}_\mathrm{loss} (1-t_\mathrm{loss}) \right. \\
&+ \left. \bar{n}_\mathrm{VTS} t_\mathrm{leak} t_\mathrm{loss} + \left(\bar{n}_\mathrm{ext} + \bar{n}_\mathrm{add} \right) t_\mathrm{loss} \right\} \!.
\end{split}
\end{equation}

Now we present our calibration protocol in three steps:

(A) $\eta_\mathrm{a}[\Delta_\mathrm{a}]$: We performed the Ramsey experiment and measured $\bar{n}_\mathrm{r}^\mathrm{eff}$ while varying $\bar{n}_\mathrm{add}$ and fixing $\bar{n}_\mathrm{VTS}=1.59$ ($T_\mathrm{VTS}=1.03\ \si{K}$).
Plotted in Fig.~\ref{fig_vts}(a) are some representative data at three different values of $\Delta_\mathrm{a}$, corresponding to the top, ridge, and floor of the right qubit-dephasing peak in Fig.~\ref{fig_tp}.
After normalizing the slopes of these $\bar{n}_\mathrm{r}^\mathrm{eff}$--$\bar{n}_\mathrm{add}$ lines with respect to the average slope of those lines at $|\Delta_\mathrm{a}| \gg \chi,\,\kappa_\mathrm{r}$, we obtain the experimental data of $1-\eta_\mathrm{a}[\Delta_\mathrm{a}]$ shown in Fig.~4(b), which is well explained by the theory in Sec.~\ref{sec_color}.

(B) $t_\mathrm{loss}$ and $t_\mathrm{leak}$: We measured $\bar{n}_\mathrm{r}^\mathrm{eff}$ while varying $\bar{n}_\mathrm{VTS}$ and fixing $\bar{n}_\mathrm{add}=0$, and fitted the data using the lines $\bar{n}_\mathrm{r}^\mathrm{eff} = \lambda \bar{n}_\mathrm{VTS} + \mu $.
As examples, data at the three selected values of $\Delta_\mathrm{a}$ are plotted and fitted in Fig.~\ref{fig_vts}(b).
The slopes of the lines are equal to
\begin{equation}
\begin{split}
\lambda[\Delta_\mathrm{a}] = \lambda_\mathrm{leak} + \lambda_\mathrm{a}[\Delta_\mathrm{a}] \xrightarrow{|\Delta_\mathrm{a}| \gg \chi,\,\kappa_\mathrm{r}}  \lambda_\mathrm{leak},
\end{split}
\end{equation}
in which
\begin{align}
\lambda_\mathrm{leak} &\coloneqq \kappa_\mathrm{r}^{-2} \kappa_\mathrm{a} \kappa_\mathrm{r,c} t_\mathrm{loss} t_\mathrm{leak},\\
\lambda_\mathrm{a}[\Delta_\mathrm{a}] &\coloneqq \kappa_\mathrm{r}^{-2} \kappa_\mathrm{a} \kappa_\mathrm{r,c} t_\mathrm{loss} \eta_\mathrm{a}[\Delta_\mathrm{a}].
\end{align}
With $\kappa_\mathrm{r}$, $\kappa_\mathrm{r,c}$, and $\kappa_\mathrm{a}$ reported in Table~S1, and $\eta_\mathrm{a}[\Delta_\mathrm{a}]$ measured in Step (A), we can deduce $t_\mathrm{loss}$ from the frequency-dependent $\lambda_\mathrm{a}[\Delta_\mathrm{a}]$, and subsequently obtain $t_\mathrm{leak}$ from the frequency-independent $\lambda_\mathrm{leak}$.

(C) $\bar{n}_\mathrm{ext}$ and $\bar{n}_\mathrm{loss}$: The vertical intercepts of the $\bar{n}_\mathrm{r}^\mathrm{eff}$--$\bar{n}_\mathrm{VTS}$ lines measured in Step (B) are equal to
\begin{equation}
\begin{split}
\mu[\Delta_\mathrm{a}] = \mu_\mathrm{loss} + \mu_\mathrm{a}[\Delta_\mathrm{a}] \xrightarrow{|\Delta_\mathrm{a}| \gg \chi,\,\kappa_\mathrm{r}} \mu_\mathrm{loss},
\end{split}
\end{equation}
in which
\begin{align}
\mu_\mathrm{loss} &\coloneqq \kappa_\mathrm{r}^{-2} \kappa_\mathrm{a} \kappa_\mathrm{r,c} \left[ \bar{n}_\mathrm{loss} (1\!-\!t_\mathrm{loss}) + \bar{n}_\mathrm{ext} t_\mathrm{loss} \right], \\
\mu_\mathrm{a}[\Delta_\mathrm{a}] & \coloneqq -\kappa_\mathrm{r}^{-2} \kappa_\mathrm{a} \kappa_\mathrm{r,c} \bar{n}_\mathrm{ext} t_\mathrm{loss} \eta_\mathrm{a} [\Delta_\mathrm{a}].
\end{align}
Therefore, with all the known parameters, we can extract $\bar{n}_\mathrm{ext}$ from the data of $\mu_\mathrm{a}[\Delta_\mathrm{a}]$, and then $\bar{n}_\mathrm{loss}$ from $\mu_\mathrm{loss}$.
Results are listed in Table~I.

\section{Classical Radiometry Data}

\begin{figure}[t]
\includegraphics[width=\columnwidth]{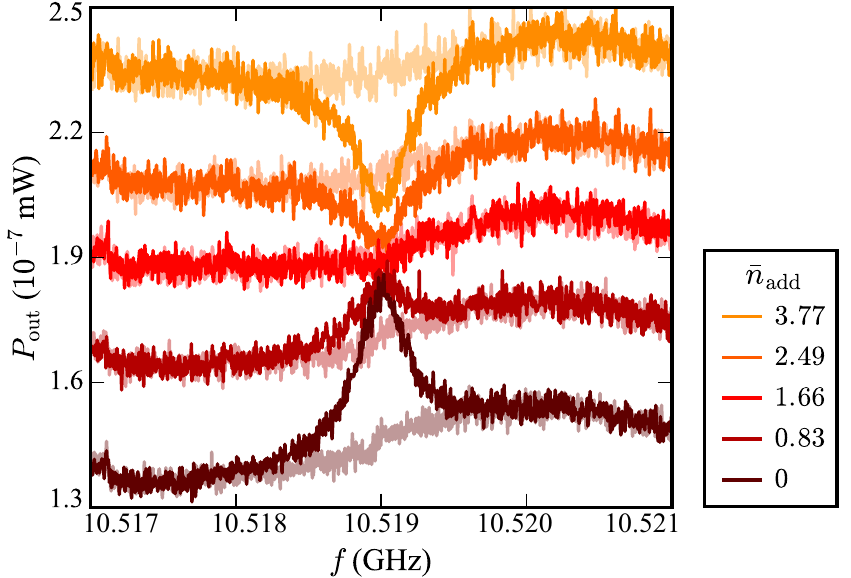}
\caption{Raw spectral density data underlying Fig.~3. The resolution bandwidth of the spectrum analyzer was set to be 1 kHz.}
\label{fig_SA}
\end{figure}

Raw spectral density data underlying Fig.~3 are shown in Fig.~\ref{fig_SA}.
The bright traces were measured with $f_\mathrm{a} = 10.519\ \si{GHz}$.
The shadowed traces were taken with the antenna resonator tuned away from the detection window.
Their differences account for the dips and peaks seen in Fig.~3, which manifest the radiative heating and cooling effects, respectively.
The floors of the spectra in Fig.~3 are acquired by averaging the distances between the shadowed background traces here in Fig.~\ref{fig_SA}, with the trace at $\bar{n}_\mathrm{add}=0$ being the reference.
With the external noise generator calibrated through the white-noise-induced dephasing experiment (see Sec.~\ref{sec_ramsey}), we can thus map $P_\mathrm{out}$ onto $\bar{n}_\mathrm{a}^\mathrm{out}$---the photon flux per unit bandwidth at the output of the antenna resonator.
The gain-saturation effect of the JPC amplifier is considered when we convert the vertical axes.

This classical radiometry method is also capable of measuring $\bar{n}_\mathrm{ext}$, and therefore, $\bar{n}_\mathrm{a}$---the thermal population of the antenna mode. To achieve this, we should measure the power spectral density at \textsf{Probe out} while varying $\bar{n}_\mathrm{VTS}$, similar to Step (B) in Sec.~\ref{sec_cal}, and calibrate the total gain of the amplifier chain. Results will be reported in an accompanying article \cite{rad_cool}.

\section{Comparison with Single-Microwave-Photon Detectors} \label{sec_det}

In the simplified linear model, a single-photon detector is characterized by two important parameters---\emph{quantum efficiency} $\eta$ and \emph{dark-count rate} $R_\mathrm{dc}$. Assuming the Poisson counting statistics, the probability of getting a ``click'' event after a time interval $\tau_\mathrm{p}$ is
\begin{align}
P_\mathrm{click} (\tau_\mathrm{p}) = 1 - e^{- (\eta R_{\bar{0}} + R_\mathrm{dc} ) \tau_\mathrm{p} - P_\mathrm{dc}^0},
\end{align}
in which $R_{\bar{0}} \tau_\mathrm{p}$ is the probability of the incoming field having nonzero photons during $\tau_\mathrm{p}$, and $P_\mathrm{dc}^0$ is the \emph{parasitic} dark-count probability.
In our case, this parameter models qubit initialization and readout errors as well as intrinsic qubit decoherence.
Therefore, the \emph{total} dark-count probability within the same detection window is given by $P_\mathrm{dc} = R_\mathrm{dc} \tau_\mathrm{p} + P_\mathrm{dc}^0$.

Define the \emph{number of clicks} during $\tau_\mathrm{p}$ as
\begin{align}
N_\mathrm{click} \coloneqq \ln\left( \frac{1}{1-P_\mathrm{click}}\right).
\end{align}
Then the quantum efficiency and the dark-count probability can be formally defined as
\begin{align}
\eta & \coloneqq \! \left. \frac{1}{\tau_\mathrm{p}} \frac{\mathrm{d} N_\mathrm{click}}{\mathrm{d} R_{\bar{0}} } \right|_{R_\mathrm{\bar{0}}=0}, \\
P_\mathrm{dc} & \coloneqq \! \left. N_\mathrm{click} \right|_{R_\mathrm{\bar{0}}=0}.
\end{align}

The Ramsey experiment combined with single-shot qubit readout can be understood from the point of view of photon detection.
After each measurement sequence, the qubit readout yielding $\ket{g}$ or $\ket{e}$ corresponds to the ``click'' or ``no click'' event of a single-photon detector.
Following this definition and the introduction to our measurement protocol in Sec.~\ref{sec_ramsey}, we find for those sequences during which the JPC switch is on for time $\tau_\mathrm{p}$, the probability of getting a ``click'' at the end of a sequence is equal to
\begin{align}
P_\mathrm{click}(\tau_\mathrm{p}) = \frac{1}{2} - A_\mathrm{on}(\tau_\mathrm{p}),
\end{align}
in which $A_\mathrm{on}$ is the amplitude of Ramsey oscillations.
If the antenna radiation is white noise, then
\begin{align}
P_\mathrm{click}(\tau_\mathrm{p}) = \frac{1}{2} - \frac{1}{2} e^{-\Gamma_\mathrm{th} \tau_\mathrm{p} - \Gamma_\mathrm{2R} (\tau-\tau_\mathrm{p})},
\end{align}
$\Gamma_\mathrm{th}$ is given by Eq.~(2) in the Main Text.

In reality, the contrast of Ramsey oscillations is smaller than unity even when $\tau=\tau_\mathrm{p}=0$.
Therefore, the exponential term in the above equation needs to be multiplied by an initial contrast factor $A_0$:
\begin{align}
P_\mathrm{click}(\tau_\mathrm{p}) = \frac{1}{2} - \frac{1}{2} A_0 e^{- \Gamma_\mathrm{th} \tau_\mathrm{p} - \Gamma_\mathrm{2R} \tau_\mathrm{w}}.
\end{align}
in which $\tau_\mathrm{w}=\tau - \tau_\mathrm{p}$ is the waiting time between the end of the JPC pump and the second $\pi/2$ qubit pulse. In our experiment, $A_0<1$ is mainly due to the readout infidelity and the imperfect qubit initialization, which cause the probability of getting a $\ket{g}$ or $\ket{e}$ outcome when $\tau=0$ to be
\begin{align}
P_g^0 &= P_g^\mathrm{ini} P_\mathrm{r}(g|g) + P_e^\mathrm{ini} P_\mathrm{r}(g|e), \\
P_e^0 &= P_e^\mathrm{ini} P_\mathrm{r}(e|e) + P_g^\mathrm{ini} P_\mathrm{r}(e|g).
\end{align}
Here $P_e^\mathrm{ini} = 1-P_g^\mathrm{ini} = 0.03$ denotes the qubit excited-state population after its initialization; $P_\mathrm{r}(e|g) = 1 - P_\mathrm{r}(g|g) = 0.01$ and $P_\mathrm{r}(g|e) = 1 - P_\mathrm{r}(e|e) = 0.04$ denote the qubit readout infidelity.
These numbers yield $P_e^0=1-P_g^0=0.0385$, and $A_0 = P_g^0 - P_e^0 = 0.923$.

In the limit of small incoming photon flux,
\begin{equation} \label{eq_white}
\Gamma_\mathrm{th} \xrightarrow{\bar{n}_\mathrm{r}^\mathrm{th} \ll 1} \frac{\chi^2 \kappa_\mathrm{r} \bar{n}_\mathrm{r}^\mathrm{th}}{\chi^2 + \kappa_\mathrm{r}^2 } = \frac{\chi^2 R_{\bar{0}}}{\chi^2 + \kappa_\mathrm{r}^2 }.
\end{equation}
Therefore, we obtain the quantum efficiency and the dark-count probability of our cQED detector based on the Ramsey sequence
\begin{align}
\eta &= \frac{A_0 e^{- \Gamma_\mathrm{2R} \tau_\mathrm{w}}}{1 + A_0 e^{- \Gamma_\mathrm{2R} \tau_\mathrm{w}}} \frac{\chi^2} {\chi^2 + \kappa_\mathrm{r}^2  }, \label{eq_eta} \\
P_\mathrm{dc} &= \ln\left( \frac{2}{1 + A_0 e^{ - \Gamma_\mathrm{2R} \tau_\mathrm{w} } }\right) \label{eq_dc}.
\end{align}
Plugging in the experimental parameters ($\tau_\mathrm{p}=1.08\ \si{\us}$ and $\tau=2.08\ \si{\us}$), we get $\eta=0.44$ and $P_\mathrm{dc} = 0.059$, which are comparable to the state-of-the-art single-microwave-photon detectors reported in recent years \cite{Inomata2016,Kono2018,Lescanne2019,Narla2016,Campagne-Ibarcq2018,Besse2018}.

\begin{figure}[t]
\includegraphics[width=\columnwidth]{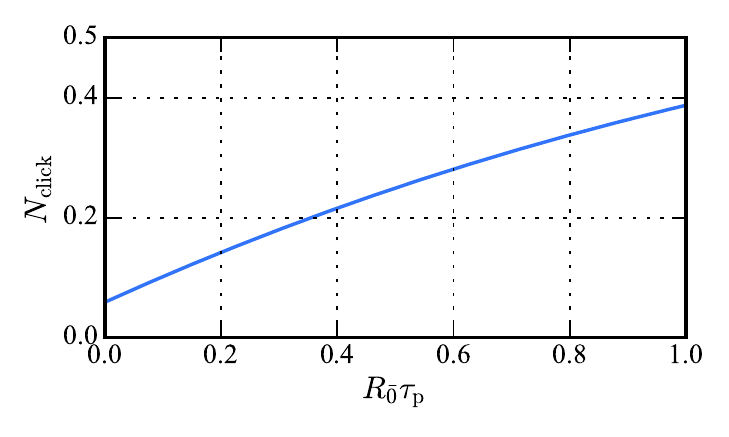}
\caption{``Number of clicks'' of the Ramsey-sequence-based qubit radiometer viewed as a photon detector. Experimental parameters are taken from Fig.~4 in the Main Text.}
\label{fig_det_eff}
\end{figure}

In Fig.~\ref{fig_det_eff}, we plot $N_\mathrm{click}$ as a function of $R_{\bar{0}} \tau_\mathrm{p}$ using the same experimental parameters as Fig.~4 in the Main Text.
The quantum efficiency is shown as the slope at $R_{\bar{0}}\rightarrow 0$.
Note that according to Eq.~(\ref{eq_eta}), the efficiency is limited to $\eta < 0.5$.
This factor-of-two sacrifice arises because a completely dephased qubit has equal probabilities of being in $\ket{g}$ and $\ket{e}$ after a Ramsey sequence.
The dark-count probability is represented by the vertical intercept.
Here we remark that the $P_\mathrm{dc} = 0.059$ reported in the present experiment is limited by the qubit-state initialization and the single-shot readout as well as the finite qubit $T_2$, all of which can be improved with available experimental techniques.
For instance, measurement-based qubit initialization plus a readout with fidelity 0.99 \cite{Narla2017} yields $P_\mathrm{dc} = 0.026$ with the present value of $T_\mathrm{2R}$.
Extending $T_\mathrm{2R}$ to 100 \si{\us} \cite{Rigetti2012}, one can then expect a dark-count probability $P_\mathrm{dc} = 0.010$, which is close to the record value reported in Ref.~\cite{Lescanne2019}.
In addition, the signal and pump of our qubit-dephasing radiometer require no pulse shaping, which results in simpler operations and more flexibility in practical tasks.

The above comparison presents our qubit-dephasing radiometer as another candidate setup for microwave-cavity-based axion probes.
As has been reasoned in Refs.~\cite{Lamoreaux2013} and \cite{Zhong2018}, single-photon detectors show advantages over quantum-limited linear amplifiers in search of dark-matter axions at high frequencies and low temperatures.
These advantages are also shared by our qubit radiometer, which essentially has the similar photon-counting performance.
Furthermore, the Peccei--Quinn mechanism \cite{Peccei1977} predicts that the galactic halo dark-matter axions induce a steady-state microwave field, which resembles the antenna radiation in this work and is thus suited for the radiometry protocol based on photon-induced qubit dephasing.
Practical challenges include shielding the superconducting circuit from the strong magnetic field, reducing the residual photon population in the qubit-readout cavity \cite{Wang2019}, and improving the stabilities of qubit coherence times.

\section{Precision Analysis}

In this section, we will compare the precisions of radiometer models based on three different physical principles---linear amplification, photon counting, and photon-induced qubit dephasing exploited in this article---given a certain detection bandwidth and integration time.
The radiator is chosen to be a broadband thermal source.
The advantage of our quantum radiometer over classical ones is quantitatively shown.

In the classical regime ($hf \ll k_\mathrm{B} T$), the precision of a total-power radiometer based on linear amplification plus square-law detection is given by \cite{Hersman1981}
\begin{align}
\frac{\delta T_\mathrm{lin}}{T_\mathrm{sys}^\mathrm{lin}} = \sqrt{\frac{2\pi}{B \tau_\mathrm{int}} + \left(\frac{\delta G}{G} \right)^2},
\end{align}
in which $T_\mathrm{sys}^\mathrm{lin}$ is the system noise temperature; $\tau_\mathrm{int}$ is the integration time; $B$ is the detection bandwidth in the unit of rad/s; $G$ and $\delta G$ are the receiver gain and its fluctuation.
The precision of a Dicke radiometer has the similar expression but a different prefactor depending on the specific type of modulation \cite{Wait1967}.
In the quantum regime, if the gain fluctuations are neglected, the above equation can be rewritten using average photon numbers,
\begin{align}\label{eq_rad_precision}
\frac{\delta \bar{n}_\mathrm{lin}}{\bar{n}_\mathrm{sys}^\mathrm{lin}} = \sqrt{\frac{2\pi}{B \tau_\mathrm{int}}}.
\end{align}
Here the precision $\delta \bar{n}_\mathrm{lin}$ is defined as \cite{Wait1967}
\begin{align} \label{eq_precision}
\delta \bar{n}_\mathrm{lin} \coloneqq \frac{\sigma_I}{\mathrm{d}\bar{I} / \mathrm{d}\bar{n}} ,
\end{align}
in which $\bar{I}$ and $\sigma_I$ are the average and the standard deviation of the detector output signal, and $\bar{n}$ is the thermal population of the source under detection.
For an ideal quantum-limited phase-preserving linear amplifier, $\bar{n}_\mathrm{sys}^\mathrm{lin*} = 1$, which comprises both the amplified vacuum fluctuations accompanying the input signal---1/2---and the minimum possible noise added by the amplifier---another 1/2.

Now consider a single-photon detector with a quantum efficiency $\eta$ and a dark-count probability $P_\mathrm{dc}$ within a detection time window $\tau_\mathrm{p}$.
In the absence of input photons, the probability of getting a ``click'' within $\tau_\mathrm{p}$ is $P_\mathrm{dc}$, with the standard deviation after $N$ counts being
\begin{align}
\sigma_I = \sqrt{\frac{P_\mathrm{dc}(1-P_\mathrm{dc})}{N}}
\end{align}
according to the binomial-distribution formula.
Given a small incoming photon flux that results in a probability $\mathrm{d} P_{\bar{0}}$ of having nonzero photons at the detector input, the probability of getting a ``click'' within $\tau_\mathrm{p}$ is then $P_\mathrm{dc} + \eta\, \mathrm{d} P_{\bar{0}}$.
The detector ``signal'' in this situation is thus
\begin{align}
\mathrm{d} \bar{I} = (P_\mathrm{dc} + \eta\, \mathrm{d} P_{\bar{0}})-P_\mathrm{dc} = \eta\, \mathrm{d} P_{\bar{0}}.
\end{align}
Following Eq.~(\ref{eq_precision}), the precision $\delta P_{\bar{0}}$ is given by
\begin{align}
\delta P_{\bar{0}} = \frac{\sigma_I}{\mathrm{d}\bar{I} / \mathrm{d} P_{\bar{0}} } = \frac{1}{\eta} \sqrt{\frac{P_\mathrm{dc}(1-P_\mathrm{dc})}{N}}.
\end{align}

In the previous section, we have demonstrated that a qubit-dephasing radiometer can be viewed as a single-photon counter with an effective quantum efficiency and a dark-count rate [Eqs.~(\ref{eq_eta}) and (\ref{eq_dc})].
For the qubit-dephasing radiometer introduced in our experiment, we have
\begin{align}
\mathrm{d} P_{\bar{0}} = \tau_\mathrm{p}\, \mathrm{d} R_{\bar{0}} = \kappa_\mathrm{r} \tau_\mathrm{p}\, \mathrm{d} \bar{n}_\mathrm{r}^\mathrm{th}.
\end{align}
Then the detection precision of $\bar{n}_\mathrm{r}^\mathrm{th}$ is
\begin{align} \label{eq_dec_precision}
\delta \bar{n}_\mathrm{qu} = \frac{\sigma_I}{\mathrm{d}\bar{I} / \mathrm{d} \bar{n}_\mathrm{r}^\mathrm{th} } = \frac{1}{ \eta \kappa_\mathrm{r} \tau_\mathrm{p} } \sqrt{\frac{P_\mathrm{dc}(1-P_\mathrm{dc})}{N}}.
\end{align}
Relating Eq.~(\ref{eq_rad_precision}) to Eq.~(\ref{eq_dec_precision}), we find the condition for the linear-amplifier-based and the qubit-dephasing-based radiometers to have the same precision, namely, $\delta \bar{n}_\mathrm{lin} = \delta \bar{n}_\mathrm{qu}$:
\begin{align}
\frac{(\bar{n}^\mathrm{lin}_\mathrm{sys})^2}{B/2\pi} = \frac{P_\mathrm{dc}(1-P_\mathrm{dc})}{\eta^2 \kappa_\mathrm{r}^2 \tau_\mathrm{p}},
\end{align}
in which we have assigned $N\tau_\mathrm{p}$ to be the effective integration time $\tau_\mathrm{int}$ for the dephasing radiometer.
Therefore, for the two radiometers to have the same precision, the linear-amplifier system should have a bandwidth
\begin{align}
B = \frac{2\pi \kappa_\mathrm{r} \tau_\mathrm{p}(\bar{n}^\mathrm{lin}_\mathrm{sys})^2 \eta^2 }{P_\mathrm{dc}(1-P_\mathrm{dc})} \kappa_\mathrm{r}.
\end{align}
Equivalently, if the linear-amplifier system has the same bandwidth as the qubit-readout cavity, namely, $B=\kappa_\mathrm{r}$, then the qubit-dephasing radiometer outperforms by a factor of
\begin{align}
\frac{\delta \bar{n}_\mathrm{lin}}{\delta \bar{n}_\mathrm{qu}} = \bar{n}_\mathrm{sys}^\mathrm{lin} \eta \sqrt{\frac{2\pi \kappa_\mathrm{r} \tau_\mathrm{p}}{P_\mathrm{dc}(1-P_\mathrm{dc})}}
\end{align}
in terms of detection precision.
Considering $\bar{n}_\mathrm{sys}^\mathrm{lin*} = 1$ for an ideal phase-preserving linear amplifier and taking in the values $\eta=0.44$ and $P_\mathrm{dc} = 0.059$ computed in Sec.~\ref{sec_det}, we obtain $\delta \bar{n}_\mathrm{lin}^* / \delta \bar{n}_\mathrm{qu} = 11$.

As is shown in Fig.~4(c) of the Main Text, when the JPC pump is on, the VTS thermal leakage and the dissipation between the antenna and the cQED cavity contribute to a parasitic white thermal background of $\bar{n}^\mathrm{para}_\mathrm{r}$ photon on average in the qubit-readout cavity.
As a result, the quantum efficiency and the dark-count rate of the whole setup are given by
\begin{align}
\eta' &= \frac{A_0 e^{- \Gamma_\mathrm{2R} \tau_\mathrm{w} - \Gamma_\mathrm{th} (\bar{n}^\mathrm{para}_\mathrm{r}) \tau_\mathrm{p} }}{1 + A_0 e^{- \Gamma_\mathrm{2R} \tau_\mathrm{w} - \Gamma_\mathrm{th} (\bar{n}^\mathrm{para}_\mathrm{r}) \tau_\mathrm{p} } } \frac{\chi^2} {\chi^2 + \kappa_\mathrm{r}^2  }, \label{eq_eta2} \\
P_\mathrm{dc}' &= \ln \left[ \frac{2}{1 + A_0 e^{ - \Gamma_\mathrm{2R} \tau_\mathrm{w} - \Gamma_\mathrm{th} (\bar{n}^\mathrm{para}_\mathrm{r}) \tau_\mathrm{p} } }\right]. \label{eq_dc2}
\end{align}
In our experimental setup, we calibrated that $\bar{n}^\mathrm{para}_\mathrm{r} = \bar{n}_\mathrm{para} t_\mathrm{loss} \kappa_\mathrm{a} \kappa_\mathrm{r,c}  / \kappa_\mathrm{r}^2 = 0.035$, which yields $\eta'=0.40$, $P_\mathrm{dc}' = 0.14$, and an outperforming factor of $\delta \bar{n}_\mathrm{lin}^* / \delta \bar{n}_\mathrm{qu}' = 6.8$ compared to an ideal quantum-limited amplifier.
Using the calibrated value of $\bar{n}_\mathrm{sys}^\mathrm{lin}=1.54$ of the classical amplifier chain (\textsf{Probe out}) \cite{rad_cool}, we obtain $\delta \bar{n}_\mathrm{lin} / \delta \bar{n}_\mathrm{qu}' = 10$.

\section{System Noise}

The single-photon-detector model of the cQED module also provides an intuitive way for us to understand the system noise of our qubit-dephasing radiometer.
Rewrite Eq.~(\ref{eq_dc2}) as
\begin{align}
P_\mathrm{dc}' &= \ln \left[ \frac{2}{1 + e^{- \Gamma_\mathrm{th} (\bar{n}^\mathrm{sys}_\mathrm{r}) \tau_\mathrm{p} } }\right], \label{eq_dc3}
\end{align}
in which
\begin{align}
\frac{\chi^2 \kappa_\mathrm{r} \tau_\mathrm{p} \bar{n}_\mathrm{r}^\mathrm{sys}}{\chi^2 + \kappa_\mathrm{r}^2 } = \frac{\chi^2 \kappa_\mathrm{r} \tau_\mathrm{p} \bar{n}_\mathrm{r}^\mathrm{para}}{\chi^2 + \kappa_\mathrm{r}^2 } + \Gamma_\mathrm{2R} \tau_\mathrm{w} - \ln A_0.
\end{align}
Therefore, the system noise of our qubit radiometer, which includes the VTS thermal leakage, the dissipation between the antenna and the cQED module, and the qubit-decoherence shot noise, when referred to the antenna input, is given by
\begin{equation}
\bar{n}_\mathrm{sys} = \bar{n}^\mathrm{sys}_\mathrm{r} \frac{\kappa_\mathrm{r}^2}{\kappa_\mathrm{r,c} \kappa_\mathrm{a} t_\mathrm{loss}} \\
= \bar{n}_\mathrm{para} + \bar{n}_\mathrm{shot},
\end{equation}
in which
\begin{align}
\bar{n}_\mathrm{shot} = \frac{(\chi^2 + \kappa_\mathrm{r}^2)\kappa_\mathrm{r}}{\chi^2 \kappa_\mathrm{r,c} \kappa_\mathrm{a} t_\mathrm{loss}} \frac{\Gamma_\mathrm{2R} \tau_\mathrm{w} - \ln A_0}{\tau_\mathrm{p}}.
\end{align}
In our experiment, $\bar{n}_\mathrm{sys} = 0.25 \pm 0.02$---the value reported in Table I of the Main Text.

\bibliography{suppl_rad_bib}